\documentclass[a4paper,fleqn]{cas-sc}
\usepackage[numbers]{natbib}
\usepackage{tabularx}
\usepackage{multirow}
\usepackage{array}
\usepackage{makecell}
\usepackage{longtable}
\usepackage{float}

\def\tsc#1{\csdef{#1}{\textsc{\lowercase{#1}}\xspace}}
\tsc{WGM}
\tsc{QE}
\usepackage{fancyhdr}
\begin{document}
\let\WriteBookmarks\relax
\def\floatpagepagefraction{1}
\def\textpagefraction{.001}

\shorttitle{PC-FDON for Electric Field Reconstruction from EFISH Measurements}

\shortauthors{Zhijian Yang}

\title [mode = title]{Polarization-Conditioned Fourier-enhanced DeepONet for Electric Field Reconstruction from EFISH Measurements}




\author[1]{Zhijian Yang}[orcid=0000-0002-8287-4267]
\cormark[1]
\ead{zhijiany@nus.edu.sg}


\affiliation[1]{organization={Department of Mechanical Engineering, College of Design and Engineering, National University of Singapore},
            city={Singapore},
%          citysep={}, % Uncomment if no comma needed between city and postcode
            postcode={117575},
            country={Singapore}}
\affiliation[2]{organization={School of Electrical Engineering, Xi,an Jiaotong University},
            city={Xi'an},
%          citysep={}, % Uncomment if no comma needed between city and postcode
            postcode={710049},
            country={China}}

\author[1]{Edwin Setiadi Sugeng}[orcid=0000-0002-7175-2467]
\ead{edwin.s@u.nus.edu}

\author[1,2]{Yaqi Zhang}[orcid=0000-0003-2930-3606]
\ead{yaqi_zhang@u.nus.edu}

\author[2]{Anbang Sun}[orcid=0000-0003-1918-3110]
\ead{anbang.sun@xjtu.edu.cn}

\author[1]{Tat Loon Chng}[orcid=0000-0001-7621-1750]
\ead{tatloon@nus.edu.sg}

\cortext[1]{Zhijian Yang}



\begin{abstract}
Electric-field-induced second-harmonic generation (EFISH) is an established laser-based diagnostic for quantifying electric fields in plasmas. Yet, ensuring field accuracy remains a challenging problem given the coherent, path-integrated nature of the signal. We address this via a machine learning approach, developing Polarization-Conditioned Fourier-enhanced Deep Operator Network (PC-FDON) -- a unified operator-learning model that reconstructs field profiles from EFISH measurements across both vertical and horizontal field polarizations and various optical parameters. Its architecture incorporates three novel advances: (\romannumeral1) a Fourier-enhanced branch providing an inductive bias for the Gouy phase shift and wave-vector mismatch - two important EFISH parameters; (\romannumeral2) a polarization-conditioning branch encoding signal polarization via Feature-wise Linear Modulation (FiLM) and gated units, enabling a \textit{single} model to handle both polarizations; and (\romannumeral3) a physics-informed loss enforcing self-consistency with the governing EFISH equation. Trained on data spanning multiple function families, polarization states, and phase-mismatch values, PC-FDON achieves promising reconstruction under noise-free, incomplete, and noisy inputs, with a generalizability comparable to our previous polarization-specific model. Crucially, pointwise epistemic uncertainty estimates are provided via Monte Carlo (MC) dropout, reflecting model confidence and enabling out-of-distribution (OOD) detection through a location-dependent exceedance fraction metric. Validation is performed on realistic electrostatic electrode configurations under both polarizations and varying Rayleigh ranges, including a simulated surface dielectric barrier discharge (SDBD) where the framework correctly flags OOD inputs, and experimental data showing good agreement with simulations. The architecture introduced – spectral inductive bias, conditional modulation, and dataset-specific uncertainty – shows strong potential for broader application beyond plasma diagnostics.
\end{abstract}

% Keywords
% Each keyword is separated by \sep

\maketitle
\thispagestyle{fancy}
\fancyhf{}
\fancyfoot[R]{\thepage}

% Main text
% ============================================================================
\section{Introduction}\label{sec:intro}
% ============================================================================
%
Electric-field-induced second-harmonic generation (EFISH) is a non-intrusive optical diagnostic that has gained considerable traction for quantifying electric fields in non-equilibrium discharges~\citep{dogariuSpeciesIndependentFemtosecondLocalized2017}. Inspired by~\citet{bigioElectricFieldInducedHarmonic1975}, the EFISH technique utilizes the nonlinear response of a plasma subjected to an external electric field ($E_{\mathrm{ext}}$) and a focused (high-intensity) probe beam ($P^{(\omega)}$), which produces a coherent second-harmonic signal ($P^{(2\omega)}$) with a temporal resolution matching that of the probe pulse duration (e.g., nanosecond or picosecond). Specifically, assuming the probe beam propagates through the plasma along the $z$-axis, the second-harmonic signal $P^{(2\omega)}$ induced by the interaction between $E_{\mathrm{ext}}(z)$ and the optical field scales quadratically with the applied field through a \textit{line-of-sight} integral involving the third-order gas hyperpolarizability $\alpha^{(3)}$, gas number density $N$, Rayleigh length $z_{\mathrm{R}}$, and the wave-vector mismatch $\Delta k$~\citep{chngElectricFieldMeasurements2020,chngEffectElectricField2022}. A key limitation of EFISH arising from the above is that a single (point) measurement does not yield a localized field value. Rather, it returns a line-of-sight-averaged quantity, dependent upon the Gouy phase shift, wave-vector mismatch, and the entire shape of the unknown electric field (not merely the focal region) and thus their bias/uncertainties~\citep{chngElectricFieldMeasurements2020,chngEffectElectricField2022}. This renders the recovery of the electric field value from the EFISH measurements an inherently ill-posed inverse problem.

Several approaches have been raised to mitigate this problem, including the use of crossed beams, designed to avoid the line-of-sight buildup error by confining the optical interaction region~\citep{raskarSpatiallyEnhancedElectric2022,hogueSensitiveDetectionElectric2023,vorenkampSuppressionCoherentInterference2023,lepikhinIncreasingEFISHSensitivity2025a}. Another approach involves translating the laser beam or external field (profile) along the propagation axis, yielding a spatially-resolved EFISH signal $P^{(2\omega)}(z_\mathrm{o})$ corresponding to the electric field $E_{\mathrm{ext}}(z)$. This profile-profile correspondence addresses the issue of  solution ill-posedness between the EFISH signal and the electric field, and is expressed as~\citep{chngEffectElectricField2022,yangDeepLearningApproach2025a}:
\begin{equation}
    P^{(2\omega)}(z_\mathrm{o}) \propto \left[\alpha^{(3)} \cdot N \cdot P_\mathrm{o}^{(\omega)}\right]^2 \cdot \frac{1}{z_\mathrm{R}} \cdot \left| \int_{-\infty}^{\infty} \frac{E_{\mathrm{ext}}(z-z_\mathrm{o}) \cdot e^{i \cdot \Delta k \cdot z}}{1 + i\cdot\frac{z}{z_\mathrm{R}}} \, d\mathrm{z} \right|^2,
    ~\label{eq:1-Efish}
\end{equation}
where $z_\mathrm{o}$ denotes the displacement between the laser beam focus and the electric field profile.
However, recovering $E_{\mathrm{ext}}(z)$ from $P^{(2\omega)}(z_\mathrm{o})$ -- the \textit{EFISH inversion problem} -- is still mathematically challenging. Several strategies have been proposed to overcome this problem; one well-established family of methods exploits axisymmetry to reformulate the reconstruction as an inverse-Abel transform, converting the problem into a deconvolution task~\citep{guoMeasurementElectricField2025a}. Alternative approaches tackle spatially inhomogeneous fields through (iterative) numerical optimization algorithms assisted by machine-learning (ML) to select an appropriate initial input field profile for inversion~\citep{chenMeasurementInhomogeneousElectric2024}. Such methods are in theory applicable to any profile shape, but may exhibit a pronounced sensitivity to the choice of the initial input profile and ML hyperparameters. Analogous challenges may also arise in discrete inversion schemes proposed recently, which rely on appropriate boundary constraints and initial parameters~\citep{luoDiscreteInversionMeasurement2025}. More recently, phase-resolved strategies have been explored as a means of performing the reconstruction~\citep{satoPhaseresolvedMeasurementElectricfieldinduced2025a}.

Another distinct strategy is offered by data-driven, deep learning (DL) frameworks, and obviates the need for strong prior assumptions as well as rule-based fitting procedures, while providing both strong generalizability and robustness~\citep{yangDeepLearningApproach2025a,yangInterpretableOperatorlearningModel2026,alicherifQuantitativeElectricfieldMeasurements2026}. The application of DL to the inverse EFISH problem was first demonstrated in our earlier work~\citep{yangDeepLearningApproach2025a}, and subsequently extended by~\citet{yangInterpretableOperatorlearningModel2026}: the former introduced a convolutional neural network (CNN) specifically for reconstructing electric fields drawn from a fuzzy-membership class of profiles, while the latter proposed a Decoder Deep Operator Network (DDON) capable of generalizing across a broader family of bell-shaped and double-peak distributions via operator learning. Both models have exhibited excellent predictive accuracy and noise resilience when applied to simulations and real-world applications (e.g., corona discharges and plasma-coupled flames). Notably, the DDON accepts ‘incomplete’ input profiles, guided by a sparse sampling strategy derived from integrated-gradient (IG) attribution. This substantially reduces the experimental effort required to determine both the sampling range and resolution while addressing the issue of poor signal sensitivity when acquiring data near the tails of bell-shaped profiles.

Given that the (EFISH) integral in Eq.~\eqref{eq:1-Efish} encodes not only the unknown shape of the electric field profile, but also (i) its polarization state; as well as (ii) optical interaction parameters (via the Gouy phase and phase mismatch), this study further explores the DL model's sensitivity to these latter two factors.

\subsection*{Polarization state of the EFISH signal}
Apart from~\citet{zhengPolarizationPropertiesEFISH2024}, a feature of EFISH that has received comparatively little attention especially in the context of inverse EFISH reconstruction, is its dependence on the component of field polarization being probed -- loosely referred to here as either vertical ($E_{\mathrm{ext},y}$) or horizontal ($E_{\mathrm{ext},x}$), related vectorially by $E_\mathrm{ext} = \sqrt{E_{\mathrm{ext},y}^2 + E_{\mathrm{ext},x}^2}$, where $x$ and $y$ are both perpendicular to the beam propagation direction $z$. The vertical component is typically defined as parallel to the interelectrode gap, and is therefore often the dominant contributor. As described in further detail below, since Eq.~\eqref{eq:1-Efish} is a tensor equation involving both vectorial and tensorial quantities such as the probe beam polarization $P^\omega$ and the gas polarizability $\alpha^{(3)}$, the polarization of the EFISH signal $P^{(2\omega)}$ is effectively sensitive to the electric field vector $E_\mathrm{ext}$~\citep{wardMolecularSecondThirdorder1975,boydNonlinearOptics2020}. In practice, assuming the electrode/reactor configuration remains unchanged, probing a particular component of the electric field simply involves monitoring that corresponding signal polarization, for instance through a polarizing element placed at the detection end of the optical setup~\citep{chngElectricFieldVector2020}.

When the polarization dependence is made explicit, the EFISH forward model in Eq.~\eqref{eq:1-Efish} can be rewritten for a given probe polarization. As an example, tracking the vertically polarized component of the EFISH signal for a vertically polarized probe beam, accesses only the $\alpha^{(3)}_{yyyy}$ tensor component and the $y$-component of the external field as shown below~\citep{wardMolecularSecondThirdorder1975,chngElectricFieldInduced2019}:
\begin{equation}
    P^{(2\omega)}_y(z_\mathrm{o}) \propto \left[\alpha^{(3)}_{yyyy} \cdot N \cdot P_{\mathrm{o},y}^{(\omega)}\right]^2 \cdot \frac{1}{z_\mathrm{R}} \cdot \left| \int_{-\infty}^{\infty} \frac{E_{\mathrm{ext},y}(z-z_\mathrm{o}) \cdot e^{i \cdot \Delta k \cdot z}}{1 + i\cdot\frac{z}{z_\mathrm{R}}} \, d\mathrm{z} \right|^2,
    ~\label{eq:2-Efish_polar}
\end{equation}
where the subscript $y$ for $P^{(2\omega)}$, $P_{\mathrm{o}}^{(\omega)}$, and $E_{\mathrm{ext}}$ denotes the vertically polarized quantities. Conversely, probing the horizontal (component of) signal polarization $P_x^{(2\omega)}$ activates a different tensor element of $\alpha^{(3)}$ and couples to the orthogonal electric field component $E_{\mathrm{ext},x}$, producing a distinct EFISH profile with a different amplitude and shape. (It is worth adding that the polarization of the probe laser also affects the EFISH signal, but this effect is only limited to its magnitude.)

Two important points should be noted – (i) the present model assumes that the polarization of the probe beam is linear, \textit{and} either parallel to $x$ or $y$. While in practice, non-orthogonal, linearly polarized probe beams are possible, the analysis of these cases is considerably more complex, and is excluded here to avoid introducing additional complexity~\citep{zhengPolarizationPropertiesEFISH2024}. (ii) Secondly, shapes of the respective field profiles ($E_{\mathrm{ext},y}$ versus $E_{\mathrm{ext},x}$) are often markedly different. Due to the latter, our previous DDON model was limited only to predicting the dominant component ($E_{\mathrm{ext},y}$).  As will be seen later, it is noted that the resulting shapes of the EFISH profiles arising from vertical and horizontal field polarizations can be qualitatively similar, a problem that further confounds polarization-specific (or agnostic) DL models. These considerations motivate the development of a polarization-conditioned model: a single, unified EFISH-resolved model that encodes the field polarization component as an explicit input and learns the distinct forward mappings associated with each polarization state.

\subsection*{Optical parameters influencing the EFISH signal}
Even for a given field polarization (e.g., vertical), variations in the phase-matching properties of the probe beam can introduce additional uncertainties into the measurements. As shown by~\citet{alicherifQuantitativeElectricField2026PROCI}, these uncertainties originate primarily from two sources: the refractive-index-dependent wave-vector mismatch $\Delta k$, and to a lesser extent, the Rayleigh length $z_\mathrm{R}$ of the focused beam. Because both parameters enter the forward model in Eq.~\eqref{eq:2-Efish_polar} as multiplicative and phase-modulating factors, their variation alters the kernel of the integral operator and, consequently, the shape of the measured EFISH profile for a given field distribution. A robust reconstruction model must therefore treat $\Delta k$ and $z_\mathrm{R}$ as variable inputs rather than training only for a specific combination of values, enabling the model to adapt to different experimental optical setups. This motivates the development of a more robust model trained on a more general dataset that learns the forward EFISH not only for a single combination of $\Delta k$ and $z_\mathrm{R}$.

\subsection*{Uncertainty quantification}
The success with which ML/data-driven approaches can be applied to the present reconstruction task (or inverse problem) suggests that it is equally important to estimate a model’s reliability, particularly when the model is deployed on experimental configurations (or profiles) that may lie outside its training distribution~\citep{bonzaniniFoundationsMachineLearning2023,wangAleatoricEpistemicExploring2025}. In the context of EFISH inversion, quantifying predictive uncertainty serves two purposes: it flags spatial regions of the reconstructed field profile where the model is least confident, guiding experimentalists toward targeted re-measurement; and also distinguishes whether the dominant source of error is epistemic or aleatoric, indicating if additional training data or improved measurement protocols would yield the greater benefit.

Epistemic uncertainty arises from incomplete knowledge of the model parameters -- finite training data, architectural choices, and optimization stochasticity all contribute -- and is in principle reducible as more data become available. Aleatoric uncertainty, by contrast, reflects intrinsic noise in the measurements (e.g., electromagnetic interference (EMI)  and detector noise) and persists regardless of model capacity or (training) data volume. Simply put, epistemic error is attributable to model imperfections/incapabilities, while aleatoric uncertainty arises due to external sources such as experimental noise.

To capture the epistemic uncertainties within the PC-FDON framework without modifying the network architecture or training objective, we adopt Monte Carlo (MC) dropout~\citep{galDropoutBayesianApproximation2016,wangAleatoricEpistemicExploring2025}, which provides a computationally and mathematically tractable approximation to Bayesian inference in deep Gaussian processes (GP) for uncertainty estimation. As its name implies, the central idea behind this approach is to perform multiple stochastic forward passes (Monte Carlo) through the model, each time randomly deactivating (dropout) a subset of its neurons, thereby obtaining an ensemble of model predictions whose spread furnishes an estimate analogous to the concept of uncertainty.  The mathematical description, which shows that this procedure is supported by Bayesian theory, and implementation details are presented in \S\ref{subsec: MC dropout}.

Taken together, the preceding considerations motivate a unified modeling framework that simultaneously accounts for polarization state, interaction parameters, and predictive uncertainty.
In this work, we introduce \textbf{PC-FDON} (\textbf{P}olarization-\textbf{C}onditioned \textbf{F}ourier-enhanced \textbf{D}eep \textbf{O}perator \textbf{N}etwork), a new architecture that addresses all three requirements within a single model. The remainder of this paper is organized as follows. \S~\ref{subsec: data pre} presents the dataset generation procedure and training methodology. \S~\ref{subsec: DDON} and \S~\ref{subsec: PC-FDON} describe the architecture of DDON and PC-FDON in detail, followed by the training details (\S~\ref{subsec: training}) and framework for assessing the model uncertainty (\S~\ref{subsec: MC dropout}). \S~\ref{sec:results} reports results on synthetic and experimental benchmarks.

% ============================================================================
\section{Methodology}~\label{sec:method}
% ============================================================================
\subsection{Data preparation}~\label{subsec: data pre}
The training dataset follows the workflow established in our previous studies~\citep{yangDeepLearningApproach2025a,yangInterpretableOperatorlearningModel2026}, spanning multiple function families to broaden model coverage and 'stress-test' generalization across families with markedly different shapes. For clarity, we briefly recap the EFISH forward formulation and note several assumptions that guide the data generation.

We adopt the EFISH forward formulation Eq.~\eqref{eq:1-Efish}, rewritten in dimensionless form as:
\begin{equation}
    P^{(2\omega)} \propto \left\{\alpha^{(3)} \cdot N \cdot P_\mathrm{o}^{(\omega)} \right\}^2\cdot E_\mathrm{o}^2 \cdot z_\mathrm{R} \cdot \left| \int_{-\infty}^\infty \frac{E_\mathrm{ext}^\prime (z^\prime-z_{\mathrm{o}}^\prime)\cdot e^{iuz^\prime}}{1+i\cdot z^\prime} \cdot dz^\prime \right|^2,
    \label{eq:3-EFISH_norm}
\end{equation}
where
\begin{equation}
E_{\mathrm{ext}}^\prime (z^\prime)= \frac{E_{\mathrm{ext}} (z^\prime)}{E_\mathrm{o}} , z^\prime= \frac{z}{z_\mathrm{R}} , z_{\mathrm{o}}^\prime= \frac{z_{\mathrm{o}}}{z_\mathrm{R}} , u=\Delta k \cdot z_\mathrm{R}.
~\label{eq:4-para_EFISH_norm}
\end{equation}
Here, the normalized field $E_{\mathrm{ext}}^{\prime}(z^{\prime})$ encodes only the shape of the profile, with the peak amplitude $E_{\mathrm{o}}$ factored out. The remaining quantities -- $z_{\mathrm{R}}$, $\Delta k$, and the prefactors in braces -- are treated either as constants along the beam propagation axis, known \textit{a priori} or readily quantified from calibration~\citep{yangInterpretableOperatorlearningModel2026,alicherifQuantitativeElectricfieldMeasurements2026}. In practice, a calibration curve suffices to recover $E_{\mathrm{o}}$, after which $E_{\mathrm{ext}}^{\prime}$ can be rescaled to obtain the absolute field. The training dataset comprises an equal mixture of vertical- and horizontal-polarization cases (viz., profiles), described below, with each subset receiving equal weightage in the final training distribution.\par

\subsubsection{Vertical polarization subset, $E_{\mathrm{ext},y}^\prime$}
For the vertical polarization subset, we follow the data-generation procedure of~\citet{yangInterpretableOperatorlearningModel2026}, employing two independent function families for the normalized field $E_{\mathrm{ext}}^{\prime}$: the Fuzzy family and the Voigt family. The latter is constructed as a convolution of Lorentzian and Gaussian profiles, yielding shapes that are distinct from those of the Fuzzy family~\citep{idaExtendedPseudoVoigtFunction2000}. This combined dataset has demonstrated strong performance for training a vertical-polarization-specific DDON with promising predictive capability. For simplicity, the dataset comprises purely symmetric profiles and is therefore expected to be applicable only to symmetric field distributions. To assess whether this assumption holds for a given reconstruction task in real-world applications, we have previously proposed a symmetry index ($\mathcal{SI}$) that quantifies the degree of asymmetry in the input EFISH profile~\citep{yangInterpretableOperatorlearningModel2026}.

Following our prior studies, the Fuzzy family is parameterized by $a \in [2,10]$, $b \in [1,3]$, and $c \in [0,5]$ (all normalized by $z_{\mathrm{R}}$), where $a$ controls the half-width at half-maximum (HWHM), $b$ adjusts the profile steepness, and $c$ sets the dip depth or peak separation. For the Voigt family, the profile is implemented via a standard pseudo-Voigt approximation - a linear combination of a Gaussian and Lorentzian component. The Gaussian width is selected as $\sigma_\mathrm{G} \in (0,10]$ and the Lorentzian half-width as $\gamma_\mathrm{L} \in (0,10]$, while $b$ and $c$ retain their same connotations and ranges as in the Fuzzy family. Because the profile shape is the key quantity governing the \textit{EFISH inversion} rather than its absolute value, all obtained $E_{\mathrm{ext},y}$ are normalized by its peak value, yielding $E_{\mathrm{ext},y}^\prime = E_{\mathrm{ext},y}/\max{\big(E_{\mathrm{ext},y}\big)}$ as indicated in Eq.~\eqref{eq:4-para_EFISH_norm}. This parameterization has been shown to span a broad continuum of physically relevant profile shapes~\citep{yangInterpretableOperatorlearningModel2026}. The polarization label for this vertical subset is denoted as $\psi = 0$.\par

\subsubsection{Horizontal polarization subset, $E_{\mathrm{ext},x}^\prime$}
For the horizontal polarization subset, we derive the $x$-component of the electric field (along the $z$-axis) from the analytical solution for a two-cylinder electrode based on~\citet{bigioElectricFieldInducedHarmonic1975}, which produces a characteristic bipolar (antisymmetric) distribution:
\begin{equation}
    E_{\mathrm{ext},x}(z) = \frac{1}{\sqrt{z^2+y^2} \cdot \operatorname{acosh}(\frac{l}{2r})} \times (\frac{z} {\sqrt{z^2+y^2}} - \frac{z}{\sqrt{z^2+(l-y)^2}}),
    \label{eq:Ex}
\end{equation}
where $x$ and $y$ denote the vertical and horizontal polarization axes, respectively, $l$ is the gap between the center of the two cylinders, and $r$ is the radius of each cylindrical electrode (cross-section). The sampling ranges are $l/z_{\mathrm{R}} \in [10,30]$ and $r/z_{\mathrm{R}} \in [1,5]$. For simplicity, we restrict $y \geq l/2$ so that a positive peak appears at $z > 0$.

To improve the generalization of the horizontal subset beyond the two-cylinder geometry, we introduce a parametric approximation that can consistently well-approximate the bipolar distribution:
\begin{equation}
E_{\mathrm{ext},x}(z) = \left[z \cdot \exp( -\left| \dfrac{z}{a \cdot c^{1/c}} \right|^c )\right]/\left[a \cdot \exp\!\left( -\dfrac{1}{c} \right)\right],
\end{equation}
where $a \in (0,10]$ generally controls the peak separation and $c \in (0,10]$ governs the decay rate, both normalized by $z_{\mathrm{R}}$. As with the vertical subset, all $E_{\mathrm{ext},x}$ profiles are normalized by their absolute peak values, yielding $E_{\mathrm{ext},x}^\prime = E_{\mathrm{ext},x}/\max{\big(\big|E_{\mathrm{ext},x}\big|\big)}$. The polarization label for the horizontal subset is denoted as $\psi = 1$.\par

\subsubsection{Optical interaction parameters and forward EFISH}
As indicated in Eq.~\eqref{eq:3-EFISH_norm}, each normalized field profile -- either $E_{\mathrm{ext},x}^{\prime}$ or $E_{\mathrm{ext},y}^{\prime}$ -- is inserted into the forward integral with an axial displacement $z_{\mathrm{o}}^{\prime}$ along the beam propagation direction, representing the translation of the applied electric field relative to the probe beam focus. An identical scanning procedure would be carried out in a typical experiment. The dimensionless phase-mismatch parameter is sampled from a discrete set $u \in \{-0.01,-0.068,-0.1,-0.2,-0.291,-0.725,-1\}$, covering a commonly encountered range in practical EFISH measurements~\citep{goldbergElectricFieldMeasurements2018,nakamuraElectricFieldMeasurement2024,yangInterpretableOperatorlearningModel2026,alicherifQuantitativeElectricfieldMeasurements2026} (see also Table~\ref{tab:optical_para} of Appendix~\ref{sec:append_optical_para}).

Following~\citet{yangInterpretableOperatorlearningModel2026}, the displacement range is bounded by \textit{envisaged} practical constraints -- for instance avoiding (physical) interference between the discharge geometry and optical components such as the focusing lens -- and we set $z_{\mathrm{o}}^{\prime} \in [-50,50]$ (for simplicity, $z^{\prime} \equiv z_{\mathrm{o}}^{\prime}$ hereafter). For numerical stability and consistency, $z_{\mathrm{o}}^{\prime}$ is max-normalized to $[-1,1]$, and $u$ is normalized by its maximum absolute value (i.e., $u^{\prime} = u/\lvert u_{\mathrm{max}}\rvert$). Because the emphasis here is on the EFISH response to profile shape, all signals are peak-normalized herein (i.e., $P_{\mathrm{norm}}=P/P_{\max}$); in evaluating Eq.~\eqref{eq:3-EFISH_norm}. In total, we generate 955\,956 profile pairs ($E_{\mathrm{ext}}^{\prime}(z^\prime) \rightleftharpoons P^{(2\omega)}_{\mathrm{norm}}(z_{\mathrm{o}}^{\prime})$) across all families, each sampled at 109 spatial points, with the horizontal ($E_{\mathrm{ext},x}^{\prime}$) and vertical ($E_{\mathrm{ext},y}^{\prime}$) subsets equally weighted at 50\% each.

As with our previous models, we apply two augmentation strategies to enhance the prediction accuracy and robustness of the model~\citep{bonzaniniFoundationsMachineLearning2023,taylorImprovingDeepLearning2018}. First, 20\% of randomly selected input profiles are cropped: the cropping window spans 50\% of the profile length with a randomized starting index, and the cropped profiles -- comprising $\{P_{\mathrm{norm}}^{(2\omega)},\, z_{\mathrm{o}}^{\prime},\,u^{\prime},\,\psi\}$ -- are upsampled via linear interpolation to match the original input resolution before being reincorporated into the training set. This exposes the model to 'partial-view scenarios' commonly encountered in practice, such as limited scanning range or incomplete profiles. Second, controlled additive Gaussian noise is injected into 20\% of inputs in each batch using a jitter layer~\citep{holmstromUsingAdditiveNoise1992,anEffectsAddingNoise1996}: specifically, zero-mean noise with a standard deviation equal to 5\% of the input signal amplitude (i.e., $\mathrm{SNR} \approx 26$~dB). This mimics sensor noise and background fluctuations frequently observed in EFISH measurements. Together, these augmentation strategies consistently improve performance in our parametric studies, enhancing reconstruction accuracy and cross-family generalization while strengthening resilience to incomplete profiles and measurement noise.\par

\subsection{Decoder DeepONet (DDON)}~\label{subsec: DDON}
DeepONet~\citep{luDeepONetLearningNonlinear2021,luLearningNonlinearOperators2021} is an operator learning architecture expressly designed to learn mappings between function spaces, often outperforming pointwise regression approaches (e.g., conventional deep neural networks, or MLPs) that target fixed input–output mappings. Inspired by these merits of operator learning, our previous work developed a decoder deepONet for predicting the vertical electric field from a vertically polarized signal profile. The standard formulation of DDON comprises encoder-decoders and two subnetworks: (i) a branch network that encodes the discretized input function $P^{(2\omega)}_{\mathrm{norm}}(z^\prime_\mathrm{o})$ into a latent vector $h(x)$ evaluated at points $x=\left\{x_1,x_2,…,x_m \right\}$; and (ii) a trunk network that encodes the observation location(s)/parameter(s) ($z^\prime={z^\prime_1,z^\prime_2,…,z^\prime_n }$) of the output or input function into a latent vector $\tau(y)$. The two streams are combined by a dot product, yielding the operator’s action at the queried location(s), $\mathcal{G}\left[h(x), \tau(y)\right]$. In essence, the hidden operators in DDON learn a mapping that takes an input function (through $h$) and returns an output function evaluated at $y$ (through $\tau$). Then, a decoder remaps the operator outputs to the physical space at the true sampling location, yielding the electric field predictions $E_{\mathrm{ext}}^{\prime}=\mathcal{D}\bigl\{\mathcal{G}\left[h(x), \tau(y)\right]\bigl\}$. It has proven capable of predicting the electric field from an unknown EFISH profile, for both synthetic and experimental data. However, the model has been trained solely on a vertical polarization dataset with a single dimensionless phase mismatch value $u=-0.068$, and is unable to handle the horizontal polarization cases with a bipolar character as well as vertically polarized profiles far beyond (approximately $\geq u=-0.068\pm30\%$) the sole training phase mismatch value. Therefore, an enhanced PC-FDON is proposed to solve these problems.

\subsection{Polarization-conditioned Fourier-enhanced DeepONet (PC-FDON)}~\label{subsec: PC-FDON}
\begin{figure*}
    \centering
    \includegraphics[width=0.9\linewidth]{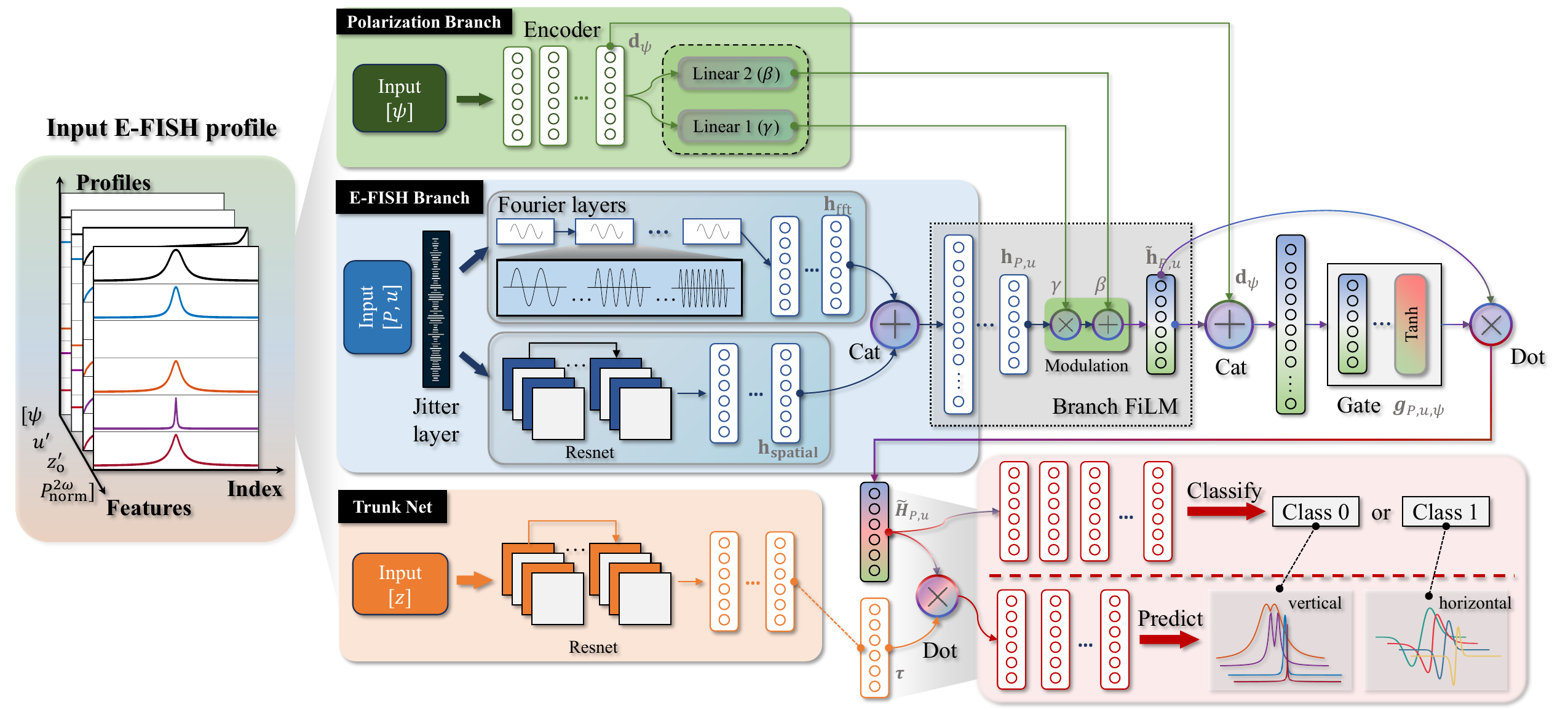}
    \caption{PC-FDON architecture for operator learning: inputs ($\left\{ P^{2\omega}_{\mathrm{norm}}(z_\mathrm{o}^\prime), u^\prime \right\}$ for EFISH branch, $\left\{ \psi \right\}$ for polarization branch, and $\left\{ z_\mathrm{o}^\prime \right\}$ for trunk net) and outputs (electric field, $E_\mathrm{ext}^\prime(z^\prime)$ and polarization class). Further details are listed in the Table~\ref{tab:PC-FDON_configuration} in Appendix~\ref{sec:append_model}.}
    \label{fig:fig1_PC-FDON}
\end{figure*}
As discussed in \S\ref{sec:intro}, recovering the horizontal electric field profile/component is equally essential in practical plasma configurations. Since the EFISH profiles produced by horizontal and vertical field polarizations \textit{may} exhibit qualitatively similar shapes, a polarization-agnostic model risks confounding such cases and misrepresenting the true field amplitude and shape. The present PC-FDON model is designed to overcome these limitations through three targeted enhancements: Fourier-domain feature extraction, polarization-aware modulation, and auxiliary polarization classification. Crucially, since both polarization configurations share the same underlying physics (Eq.~\eqref{eq:3-EFISH_norm}), training a single unified model is more data-efficient than maintaining separate polarization-specific models and provides consistent uncertainty quantification across both measurement configurations.

The PC-FDON architecture (Fig.~\ref{fig:fig1_PC-FDON}) comprises four modules:

\textbf{\romannumeral1.\ Fourier-enhanced Branch-1 (EFISH branch).}
We augment the DDON branch network with Fourier layers -- network layers that operate in the frequency domain rather than directly on the (spatial) input profile. The motivation for this design draws on two recent successes in plasma diagnostics: the inversion of plasma emission intensity via inverse Fourier analysis~\citep{parkSpatiallyResolvedEmission2010} and the prediction of the spatiotemporal evolution of laser-produced plasma via the Fourier Neural Operator (FNO)~\citep{wuDeepLearningbasedSpatiotemporal2025a}. In a typical neural network, operations are performed entirely in the physical space (e.g., the DDON learns the spatial evolution of the EFISH signal along the beam propagation axis). A Fourier layer, by contrast, operates in the frequency domain. It first transforms the data into the frequency domain using a fast Fourier transform (FFT), applies a learned weighting to the resulting spectral coefficients, then transforms these back into physical space via an inverse FFT~\citep{liFourierNeuralOperator2021,bruntonPromisingDirectionsMachine2024,congDeeponetModifiedFNO2026a,santosHybridDeepONetSurrogates2026}. Such a spectral representation is particularly well suited to the EFISH inverse problem, as it enables the network to efficiently extract multi-scale spectral features from the EFISH signal profile -- features that are related to
the Gouy phase shift kernels and the spatial frequency content of the underlying electric field. The Fourier layers enable the network to learn these signatures directly, without requiring the inverse kernel to be derived analytically. It should be added that direct application of the Fourier transform however, is insufficient, given the modulus in Eq.~\eqref{eq:1-Efish}, which results in a loss of phase information.

In the current model, the EFISH branch comprises two parallel sub-branches (Fig.~\ref{fig:fig1_PC-FDON}). The Fourier sub-branch takes two inputs -- the normalized EFISH signal $P_{\mathrm{norm}}^{(2\omega)}(z^{\prime}_\mathrm{o})$ and phase-mismatch parameter $u^{\prime}$ -- and compresses them into a compact spectral feature vector $\mathbf{h}_{\mathrm{fft}}$, which captures the global frequency content of the signal. In parallel, the ResNet sub-branch (inherited from the DDON) processes the same EFISH signal through a series of residual blocks to produce a spatial feature vector $\mathbf{h}_{\mathrm{spatial}}$, which captures local details such as peak positions and gradient information. The two vectors are then concatenated (joined end-to-end) to form a combined feature vector $\mathbf{h}_{P,u}$. This dual-path design ensures that the model retains detailed spatial information (from the ResNet sub-branch) while simultaneously leveraging global frequency content (from the Fourier sub-branch), producing a richer representation than either path alone could provide.

\textbf{\romannumeral2.\ Polarization-conditioned Branch-2 (polarization branch), FiLM and gated unit.}
As discussed in \S\ref{sec:intro}, a given beam polarization interacts with different components of the electric field (either vertical or horizontal), producing distinct EFISH profiles for a particular underlying field distribution. However, because both polarization configurations can yield profiles belonging to the same shape classes -- bell-shaped and double-peaked -- a model cannot reliably distinguish between polarization states from the EFISH signal alone. A dedicated polarization branch is therefore introduced to resolve this ambiguity. Its role is to inform the model \emph{which} polarization configuration was used, so that it can interpret the EFISH signal accordingly, similar to how an experimentalist would apply a different calibration curve depending on the polarization component.

This branch takes the binary polarization label $\psi$ as an input ($\psi = 0$ for vertical, $\psi = 1$ for horizontal, as earlier described). Through a lightweight multilayer perceptron (MLP), the branch produces three sets of outputs via three output heads: (i) a dense embedding vector $\mathbf{d}_{\psi}$ (that encodes the polarization character so the model can process this alongside other features), (ii) a scaling vector $\boldsymbol{\gamma}$ (from Linear~1), (iii) and a shift vector $\boldsymbol{\beta}$ (from Linear~2). These latter two vectors are used to adjust the combined feature vector $\mathbf{h}_{P,u}$ (from the EFISH branch) through a method known as `Feature-wise Linear Modulation' (FiLM)~\citep{perezFiLMVisualReasoning2018}. The idea is straightforward: each element of the EFISH feature vector $\mathbf{h}_{P,u}$ is independently rescaled and shifted according to the polarization states:
\begin{equation}
    \widetilde{\mathbf{h}}_{P,u}
    = \boldsymbol{\gamma} \otimes \mathbf{h}_{P,u} + \boldsymbol{\beta},
    \label{eq:film}
\end{equation}
where $\otimes$ denotes element-wise multiplication. Intuitively, $\boldsymbol{\gamma}$ controls \emph{how much} each learned feature (from the EFISH branch) contributes to the reconstruction (amplifying informative features and suppressing irrelevant ones), while $\boldsymbol{\beta}$ adds a polarization-dependent offset. This enables the model to apply different adjustments to its features depending on which polarization was used.

The FiLM-modulated vector $\widetilde{\mathbf{h}}_{P,u}$ (Eq.~\eqref{eq:film}) is then concatenated with the dense embedding vector $\mathbf{d}_{\psi}$ from the remaining output head, which provides an additional direct pathway for polarization information to enter the subsequent processing stage. The concatenated vector is passed through a \emph{gated unit}-- a dense layer followed by a $\tanh$ activation -- that compresses the representation into a bounded range $[-1,\, 1]$, yielding the polarization-conditioned vector $\mathbf{g}_{P,u,\psi}$, see Fig.~\ref{fig:fig1_PC-FDON}. To retain information from both the gated and ungated representations, the two vectors are combined via element-wise multiplication and projected through a final dense layer, yielding $\widetilde{\boldsymbol{\mathcal{H}}}_{P,u}$.
\begin{equation}
    \widetilde{\boldsymbol{\mathcal{H}}}_{P,u}
    = \mathbf{g}_{P,u,\psi} \otimes \widetilde{\mathbf{h}}_{P,u}
    \label{eq:gate-proj}
\end{equation}
The element-wise product between the gated feature vector $\mathbf{g}_{P,u,\psi}$ and ungated vectors $\widetilde{\mathbf{h}}_{P,u}$ acts as a soft selection mechanism: when the $\tanh$ gate produces values near $\pm 1$, the corresponding features pass through largely unchanged; if the gate drives values toward zero, those features are effectively suppressed. The outcome is a single polarization-conditioned feature vector $\widetilde{\boldsymbol{\mathcal{H}}}_{P,u}$ that encodes both the EFISH profile content and the polarization context, enabling a \emph{single unified model} to handle both polarization configurations without retraining. It should be emphasized that even though the model explicitly requires the polarization state of the EFISH profile ($\psi$) as a separate input, it should not be viewed as the combination of two distinct models which have been separately trained for each polarization shape. Rather, the single model retains the same weights and hyperparameters for both polarizations, which in turn is used for uncertainty quantification. And it is this feature -- facilitated by adopting FiLM and gated units -- which allows it to better `learn' the specifics of our underlying problem, thereby supporting the notion of a `unified model'.

\textbf{\romannumeral3.\ Trunk network and electric field prediction head.}
The trunk network and decoder are inherited from the DDON architecture. The trunk maps the query coordinate $z_{\mathrm{o}}^{\prime}$ to a set of basis functions $\boldsymbol{\tau}(z_{\mathrm{o}}^{\prime})$ through a series of residual blocks followed by several dense layers~\citep{yangInterpretableOperatorlearningModel2026}. The inner product of $\boldsymbol{\tau}$ with the polarization-conditioned latent vector $\widetilde{\boldsymbol{\mathcal{H}}}_{P,u}$ produces the operator output, which is subsequently refined by a decoder MLP to yield the reconstructed field $E_{\mathrm{ext}}^{\prime}(z^{\prime})$.

\textbf{\romannumeral4.\ Polarization classification head.}
In parallel with the reconstruction pathway, $\widetilde{\boldsymbol{\mathcal{H}}}_{P,u}$ is fed into a lightweight classification head consisting of several dense layers followed by a sigmoid activation. This auxiliary head predicts the polarization label $\hat{\psi} \in [0,1]$, serving two purposes: it provides an explicit supervisory signal that encourages the latent space to encode polarization-discriminative features, and it offers a built-in indicator/diagnosis at inference time. A confident classification confirms that the model has correctly identified the probe-laser configuration, while an uncertain prediction flags potential ambiguity in the input signal.

\textbf{\romannumeral5.\ Physics-informed learning.}
The total training objective combines three terms---a data-fidelity loss, a physics-consistency loss, and a classification loss---weighted by hyperparameters $\lambda_{\mathrm{PINN}}$ and $\lambda_{\mathrm{CLS}}$:
\begin{align}
    \mathcal{L}
    &= \mathcal{L}_{\mathrm{data}}
     + \lambda_{\mathrm{PINN}}\,\mathcal{L}_{\mathrm{phys}}
     + \lambda_{\mathrm{CLS}}\,\mathcal{L}_{\mathrm{cls}}
    \notag\\[4pt]
    &= \underbrace{
    \mathrm{MSE}\left\{E_{\mathrm{ext}}^{\prime},\hat{E}_{\mathrm{ext}}^{\prime}
       \right\}}_{\mathcal{L}_{\mathrm{data}}}
     + \lambda_{\mathrm{PINN}}\cdot
       \underbrace{
       \mathrm{MSE}\left\{ P^{(2\omega)},\hat{P}^{(2\omega)}
       \right\}}_{\mathcal{L}_{\mathrm{phys}}}
     + \lambda_{\mathrm{CLS}}\cdot
       \underbrace{
       \mathrm{BCE}\left\{\psi, \hat{\psi}\right\}}_{\mathcal{L}_{\mathrm{cls}}},
    \label{eq:loss}
\end{align}
where $\hat{E}_{\mathrm{ext}}^{\prime}$ is the PC-FDON prediction, $\hat{P}^{(2\omega)}$ is the EFISH signal reconstructed from $\hat{E}_{\mathrm{ext},i}^{\prime}$ via the forward integral of Eq.~\eqref{eq:3-EFISH_norm}, and $\hat{\psi}_i$ is the predicted polarization label from the classification head. The data-fidelity term $\mathcal{L}_{\mathrm{data}}$ penalizes the mean squared error between the predicted and ground-truth electric field profiles. The physics-informed term $\mathcal{L}_{\mathrm{phys}}$ enforces self-consistency with the forward integration model by penalizing the mismatch between the measured and reconstructed EFISH signals, acting as a soft physical constraint. The classification term $\mathcal{L}_{\mathrm{cls}}$ is the binary (or sigmoid) cross-entropy (BCE) between the predicted and true polarization labels, encouraging the model to encode polarization-specific features. The weighting hyperparameters $\lambda_{\mathrm{PINN}}$ and $\lambda_{\mathrm{CLS}}$ balance the relative contributions of the three objectives; their selection is detailed in \S\ref{subsec: training}.\par

\subsection{Training details}\label{subsec: training}
The PC-FDON accepts the following group of normalized parameters including the EFISH signal profile (either horizontally or vertically polarized) $\{P^{(2\omega)}_\mathrm{norm},z_\mathrm{o}^\prime,u^\prime,\psi\}$ as an input and predicts the normalized electric field shape and corresponding polarization class $\{E_{\mathrm{ext}}^{\prime}(z^{\prime}),\psi_\mathrm{P}\}$. The full dataset of $955\,956$ profile ($1\,147\,147$ after augmentation) pairs is randomly partitioned into three subsets: 90\% for training, 5\% for validation, and 5\% for testing. The validation set supports hyperparameter tuning and checkpoint selection, while the held-out test set provides an unbiased estimate of generalization performance. Given the large size of the dataset, a 90\%-5\%-5\% split provides sufficient samples for reliable validation and testing while maximizing the data available for training.

Training is performed using the Adam optimizer with an initial learning rate of $10^{-3}$ and a reduce-on-plateau scheduler that halves the learning rate after 5 consecutive epochs without improvement in the validation loss. The batch size is set to 512. Unless
otherwise specified, Gaussian error linear units (GELUs) are used in all convolutional and dense layers. Further
details are listed in the Table~\ref{tab:PC-FDON_configuration} in Appendix~\ref{sec:append_model}. The composite loss of Eq.~\eqref{eq:loss} is minimized with the physics-informed penalization coefficient $\lambda_{\mathrm{PINN}} = 0.05$ and the classification coefficient $\lambda_{\mathrm{CLS}} = 0.5$. Early stopping terminates training if the validation loss fails to improve for 15 successive epochs, with a maximum budget of 200 epochs~\citep{bengioPracticalRecommendationsGradientbased2012,yangInterpretableOperatorlearningModel2026}. In practice, this regimen yields stable optimization, striking a balance between reconstruction accuracy and the regularization needed for reliable inversion. All models are implemented in PyTorch and trained on the National University of Singapore (NUS) Hopper high-performance computing cluster with NVIDIA H100 GPUs.\par

\subsection{Uncertainty quantification via MC dropout}~\label{subsec: MC dropout}
Uncertainty quantification of DL models has become an essential component of estimating a model's reliability~\citep{galDropoutBayesianApproximation2016,abdarReviewUncertaintyQuantification2021,wangAleatoricEpistemicExploring2025}. Standard approaches include ensemble learning methods, Bayesian neural networks, deep Gaussian processes, and MC dropout~\citep{bonzaniniFoundationsMachineLearning2023,wangAleatoricEpistemicExploring2025}. The latter approach: MC dropout, which is adopted in our work, provides an approximation to Bayesian inference in deep Gaussian processes~\citep{galDropoutBayesianApproximation2016}. The central insight is that a neural network with a dropout layer applied before every dense layer can be interpreted as a variational approximation to a specific deep GP, where the choice of activation function (e.g., ReLU or $\tanh$) determines the form of the corresponding covariance~\citep{galDropoutBayesianApproximation2016,abdarReviewUncertaintyQuantification2021}. A practical advantage of this approach is that it requires no modification to the network architecture or training procedure but only the inference procedure during testing~\citep{galDropoutBayesianApproximation2016}, and has thus been applied in uncertainty quantification for various kinds of networks such as CNN, U-Net, \textit{etc.}~\citep{tousignantPredictionDiseaseProgression2019,abdarReviewUncertaintyQuantification2021,kasimBuildingHighAccuracy2021}. During inference, $N$ stochastic forward predictions with a frozen dropout layer enabled (frozen during training) are performed for each input EFISH profile. Each pass randomly freezes a fraction of the hidden units in the dense layer, yielding a collection of $N$ reconstructed field profiles $\{E_{\mathrm{ext,P}}^{\prime\,(n)}(z^{\prime})\}_{n=1}^{N}$. The predictive mean and standard deviation are then obtained as:
\begin{equation}
    \bar{E}_{\mathrm{ext,P}}^{\prime}(z^{\prime})= \frac{1}{N}\sum_{n=1}^{N} E_{\mathrm{ext,P}}^{\prime\,(n)}(z^{\prime}),
    \qquad
    \sigma(z^{\prime}) = \sqrt{\frac{1}{N}\sum_{n=1}^{N}
    \Bigl(E_{\mathrm{ext,P}}^{\prime\,(n)}(z^{\prime})
    - \bar{E}_{\mathrm{ext,P}}^{\prime}(z^{\prime})\Bigr)^2}.
    \label{eq:mc-var}
\end{equation}
The standard deviation (STD, $\sigma(z^{\prime})$) serves as a \textit{pointwise} uncertainty estimate along the laser propagation path. A large $\sigma$ at a given location $z^{\prime}$ indicates that the reconstruction is sensitive to the activated neurons sampled, signaling epistemic uncertainty at that point. Conversely, a low $\sigma$ suggests that the model parameters are well-determined and that the residual reconstruction error is more likely dominated by aleatoric contributions from measurement noise. In this work, we use $N = 100$ stochastic forward passes, a value that provides stable standard deviation estimates while remaining computationally inexpensive.

To establish a reference scale for interpreting the uncertainty at inference time, we first evaluate the MC dropout procedure across the entire validation set and compute, at each spatial location $z_j^{\prime}$ ($j = 1, 2, \ldots, n$), the empirical distribution of pointwise STD. From this distribution we extract the location-specific percentile thresholds $\sigma_{\mathrm{val}}^{(p95)}(z_j^{\prime})$ and $\sigma_{\mathrm{val}}^{(p99)}(z_j^{\prime})$. These location-dependent baselines are defined considering that certain spatial regions (e.g., the tails of a field profile) may exhibit intrinsically higher uncertainty than others even under in-distribution conditions. We then compute the exceedance fractions $\mathcal{F}_{\mathrm{val}}^{(p95)}$ and $\mathcal{F}_{\mathrm{val}}^{(p99)}$ over the validation set itself, which serve as the in-distribution reference values for the confidence assessment defined below.

When the model is applied to an EFISH input, we compare the $\sigma(z_j^{\prime})$ at each location against the corresponding validation threshold and compute the exceedance fraction for the entire profile:
\begin{equation}
    \mathcal{F}^{(p)}
    = \frac{1}{n}\sum_{j=1}^{n}
      \mathbb{1}\!\bigl[\sigma(z_j^{\prime})
      > \sigma_{\mathrm{val}}^{(p)}(z_j^{\prime})\bigr],
    \label{eq:exceedance}
\end{equation}
where $\mathbb{1}[\cdot]$ is the indicator function and $p \in \{p95, p99\}$ denotes the chosen percentile. The exceedance fraction $\mathcal{F}^{(p)}$ quantifies the proportion of spatial points at which the model's epistemic uncertainty exceeds the $p$-th percentile of the validation baseline. A profile-level confidence assessment is then assigned by comparing its own exceedance fraction against the corresponding validation reference:
\begin{equation}
    \text{Confidence level} =
    \begin{cases}
        \text{very high (in-distribution)},
        & \text{if } \mathcal{F}^{(p95)}
          \leq \mathcal{F}_{\mathrm{val}}^{(\mathrm{median})}, \\[4pt]
        \text{high (in-distribution)},
        & \text{if } \mathcal{F}_{\mathrm{val}}^{(\mathrm{median})}
          < \mathcal{F}^{(p95)}
          \leq \mathcal{F}_{\mathrm{val}}^{(p95)}, \\[4pt]
        \text{borderline (treat with care)},
        & \text{if } \mathcal{F}_{\mathrm{val}}^{(p95)}
          < \mathcal{F}^{(p95)}
          \leq \mathcal{F}_{\mathrm{val}}^{(p99)}, \\[4pt]
        \text{low (out-of-distribution)},
        & \text{if } \mathcal{F}^{(p99)}
          > \mathcal{F}_{\mathrm{val}}^{(p99)}.
    \end{cases}
    \label{eq:confidence-levels}
\end{equation}
If the $p95$ exceedance fraction falls at or below the median of the validation exceedance distribution, the reconstruction is deemed very high-confidence, indicating that the input lies comfortably within the learned manifold. If it exceeds the validation median but remains at or below the validation $p95$ reference, the confidence is high -- the model's uncertainty is consistent with the upper range of in-distribution behavior. If the $p95$ exceedance fraction surpasses the validation $p95$ reference but stays within the validation $p99$ bound, borderline behavior is flagged, and the reconstruction should be interpreted with care. Finally, if the $p99$ exceedance fraction exceeds its validation counterpart, the reconstruction is classified as low-confidence and thus assessed as having a high probability of being out-of-distribution (OOD), signaling that the predicted field profile should be treated with caution. Furthermore, because the exceedance fraction is a global metric that aggregates over an entire spatial profile, it can be insensitive to extreme localized anomalies affecting a few spatial points. In such cases, even if the exceedance fraction indicates borderline confidence, a reconstruction with $\sigma(z^\prime_j)$ significantly exceeding the $p99$ reference at specific locations should be treated with greater caution or even as OOD.\par

% ============================================================================
\section{Results and discussion}~\label{sec:results}
% ============================================================================
%
\subsection{Polarization-conditioned predictions}~\label{subsec:results_polar}
\begin{figure*}
    \centering
    \includegraphics[width=1\linewidth]{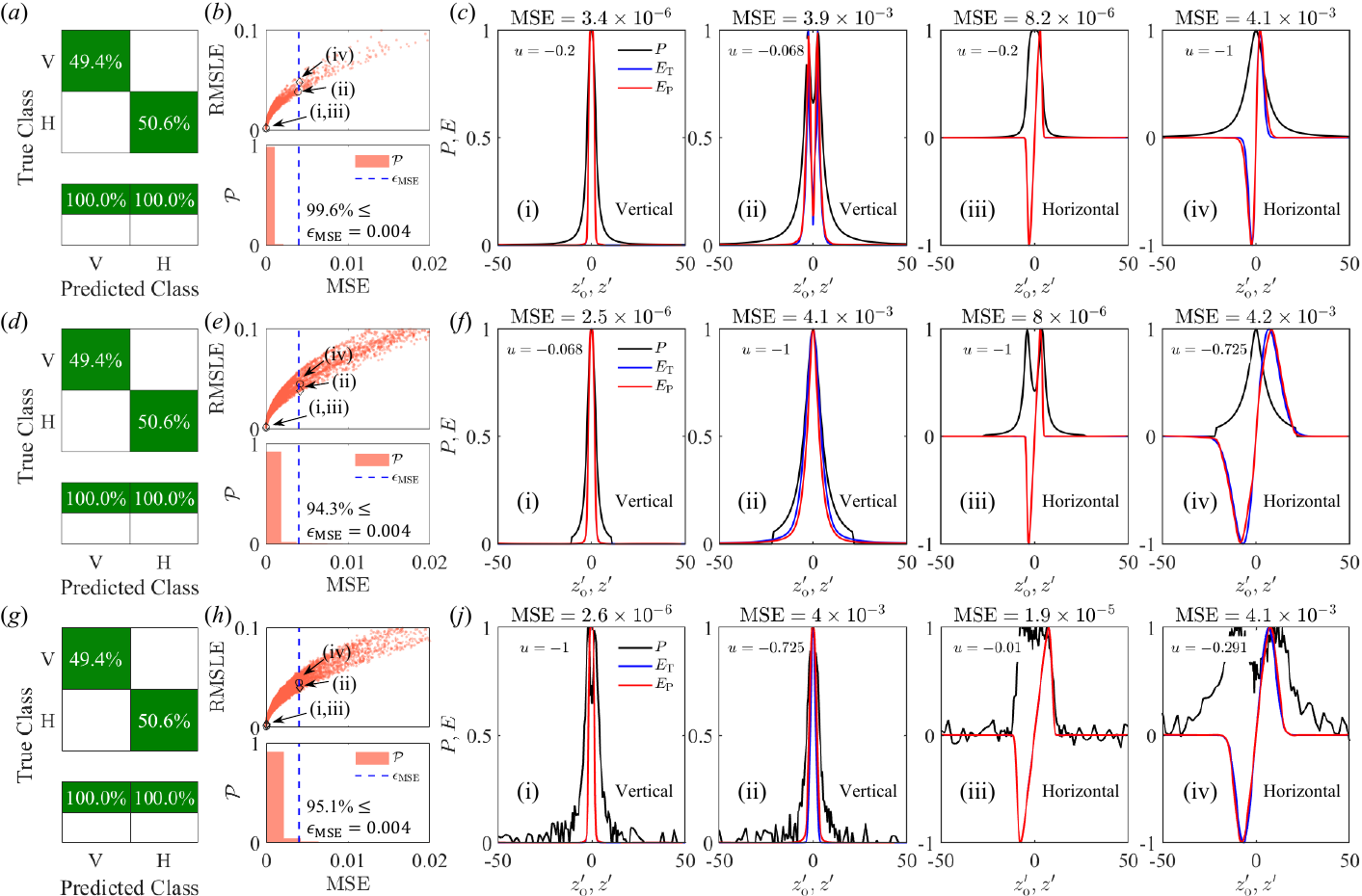}
    \caption{Polarization-conditioned reconstruction performance of PC-FDON under three input conditions: (a--c)~noise-free, (d--f)~incomplete, and (g--j)~noisy ($\mathrm{SNR} = 20$~dB). Predictions are shown for both the vertical ($E_{y}^{\prime}$, panels~i--ii) and horizontal ($E_{x}^{\prime}$, panels~iii--iv) polarization states. For each condition, the first column (a,\,d,\,g) displays the polarization classification accuracy via a confusion matrix, the second column (b,\,e,\,h) shows the prediction error distribution in the MSE--RMSLE space, and the remaining columns present four representative reconstructions overlaid on the ground truth $E_\mathrm{T}$.}
    \label{fig:fig2_Results_PC-FDON_robust}
\end{figure*}
The polarization-conditioned predictions of PC-FDON are presented in Fig.~\ref{fig:fig2_Results_PC-FDON_robust}, demonstrating the model's strong reconstruction capability under three progressively challenging input conditions: (a--c)~noise-free, (d--f)~incomplete, and (g--j)~noisy ($\mathrm{SNR} = 20$~dB) EFISH profiles. Across all three scenarios, the model accurately recovers both the vertical and horizontal electric field components (Fig.~\ref{fig:fig2_Results_PC-FDON_robust}a,d,g) from input EFISH signals whose bell-shaped envelopes are qualitatively similar -- and in many cases nearly indistinguishable -- despite originating from different polarization configurations.

To visualize profile-specific performance, a diagnostic diagram is constructed by plotting the root-mean-square logarithmic error (RMSLE) against the MSE for each test sample, where the former is more sensitive to errors in low-amplitude regions while the latter to errors in high-amplitude regions near the profile peak. The vast majority of test samples cluster in the lower-left corner of this space, indicating simultaneously low errors across both metrics, while a small number of outliers are spread toward $\mathrm{MSE} \gtrsim 4 \times 10^{-3}$. Representative predictions with $\mathrm{MSE} \lesssim 4 \times 10^{-3}$, selected to span the range of typical reconstruction quality, are also displayed; these examples show close agreement with the ground truth across the full spatial domain. Using $\epsilon_{\mathrm{MSE}} = 4 \times 10^{-3}$ as a threshold, the fractions of predictions satisfying $\mathrm{MSE} \leq \epsilon_{\mathrm{MSE}}$ are 99.6\%, 94.3\%, and 95.1\% for the noise-free, incomplete, and noisy conditions (Fig.~\ref{fig:fig2_Results_PC-FDON_robust}b,e,h), respectively. Notably, the model demonstrates robust performance in two additional respects. First, accurate reconstructions are maintained for phase-mismatch values spanning the full training range $u \in [-0.01, -1]$, confirming that the model successfully generalizes over the interaction parameters encoded as variable inputs. Second, the model tolerates incomplete input profiles in which the EFISH signal is zeroed beyond $z^\prime \geq 4.2 \times \mathrm{FWHM}$, indicating that it has inherited a similar key feature range identified earlier for the DDON, used to define a general sampling criterion previously in~\citet{yangInterpretableOperatorlearningModel2026} based on IG.\par

\subsection{Generalization across noisy function spaces}
\begin{figure*}
    \centering
    \includegraphics[width=1\linewidth]{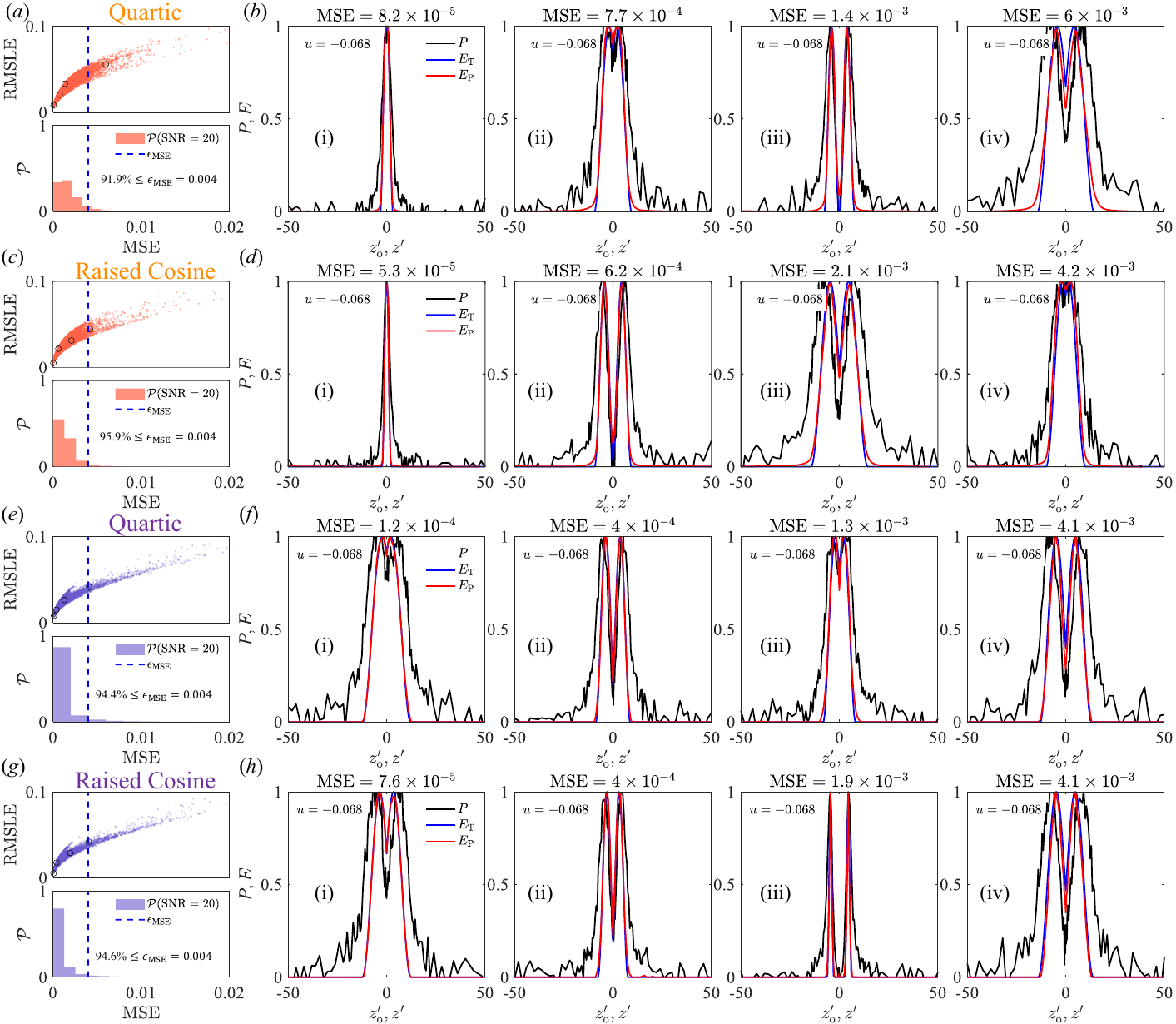}
    \caption{Generalization performance of PC-FDON (orange) and DDON ~\citep{yangInterpretableOperatorlearningModel2026} (purple) on two unseen function families under noisy conditions ($\mathrm{SNR} = 20$~dB): (a,b,e,f)~Quartic and (c,d,g,h)~Raised Cosine. Panels (a--d) show PC-FDON predictions; panels (e--h) show DDON predictions. Both models are evaluated only for the vertically polarized field component, as the earlier DDON model was designed for predicting only that specific (vertical) polarization. Each function space contains 9\,191 test profiles.}
    \label{fig:fig3_Results_PC-FDON_generlization}
\end{figure*}

To assess the function-space generalizability of PC-FDON and compare it with the DDON baseline, we evaluate both models on unseen profiles drawn from two function families not represented in the training set: (i) Quartic and (ii) Raised Cosine. To ensure a fair comparison, both models are evaluated on the same dataset, identical to that used in~\citet{yangInterpretableOperatorlearningModel2026}. These families span a broad spectrum of bell-shaped and double-peak profiles, providing a rigorous test of model generalizability across distinct function classes. Rather than using noise-free inputs, we conduct the comparison under noisy conditions ($\mathrm{SNR} = 20$~dB), which offer a more demanding assessment of each model's robustness.

As shown in Fig.~\ref{fig:fig3_Results_PC-FDON_generlization}, PC-FDON achieves comparable performance to DDON across both function spaces (each containing 9\,191 profiles), with both models exhibiting low prediction errors. The DDON holds a slight edge, manifested as a left-shifted MSE distribution with a marginally higher concentration of low-error predictions. Under more extreme noise levels (e.g., $\mathrm{SNR} = 15$~dB; not shown), the performance gap widens, with PC-FDON exhibiting a larger degradation in reconstruction accuracy. This outcome is not unexpected: the DDON is a model trained exclusively on the vertical polarization state with a single phase-mismatch value, whereas PC-FDON must allocate its representational capacity across both polarization configurations and various phase-mismatch parameters. In particular, although the total training set for PC-FDON is larger (955\,956 pairs vs.\ 703\,682 pairs for the DDON), the vertical polarization subset constitutes only approximately half of that total, meaning the effective number of vertical-polarization training examples available to PC-FDON is much less than that of the DDON. We anticipate that enlarging the training corpus, particularly the vertical polarization subset, and extending the training schedule will close this gap.\par

\subsection{Practical applications and uncertainty quantification}
In this section, we apply PC-FDON to several realistic electrostatic configurations to reconstruct unknown electric field distributions from both simulated and experimental data. Each case is accompanied by the MC dropout uncertainty analysis described in \S\ref{subsec: MC dropout}.

\subsubsection{Surface Dielectric Barrier Discharge (sDBD) simulations}~\label{subsubsec: SDBD}
\begin{figure*}
    \centering
    \includegraphics[width=1\linewidth]{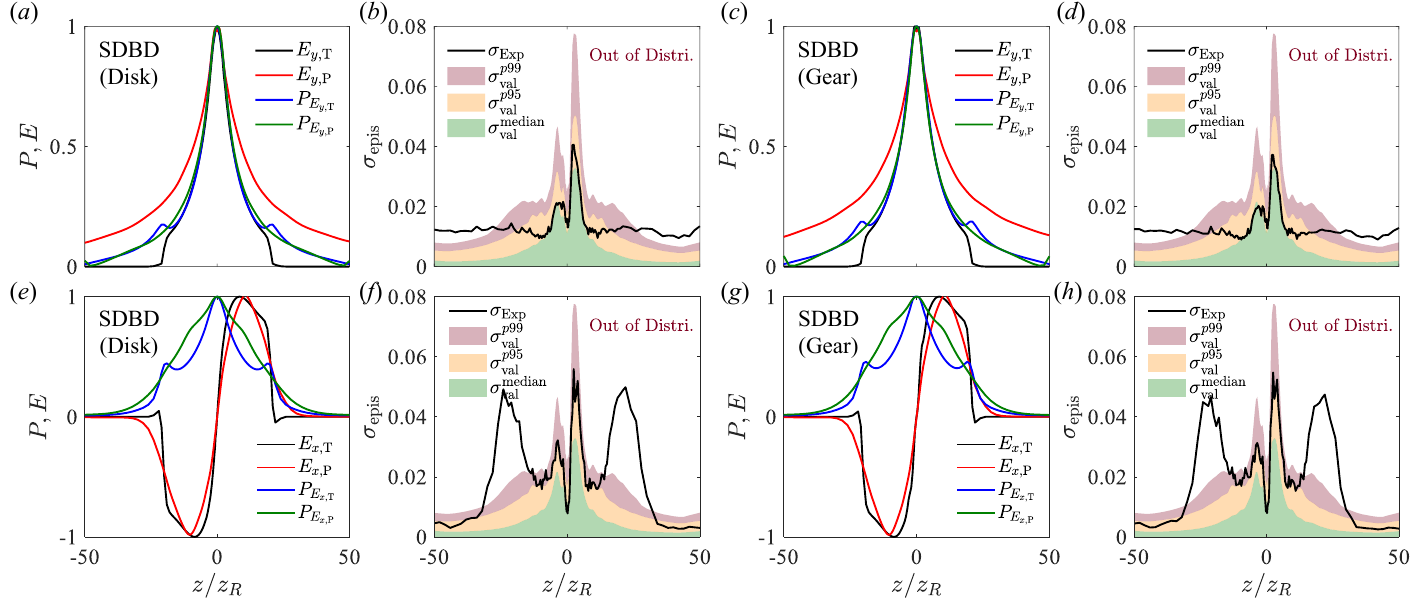}
    \caption{Predicted electric field components (horizontal: $E_{x,\mathrm{P}}$; vertical: $E_{y,\mathrm{P}}$) and associated uncertainty quantification for (a,b,e,f)~disk and (c,d,g,h)~gear SDBD configurations. Panels (a,e) and (c,g) show the reconstructed field profiles benchmark against the simulation ground truth; panels (b,f) and (d,h) show the pointwise standard deviation $\sigma(z^{\prime})$ from 100 MC dropout passes compared against the validation percentile thresholds $\sigma_{\mathrm{val}}^{(p95)}$ and $\sigma_{\mathrm{val}}^{(p99)}$. Both configurations are flagged as out-of-distribution (OOD) by the exceedance criterion of Eq.~\eqref{eq:confidence-levels}.}
    \label{fig:fig4_Sim_dbd}
\end{figure*}
A convenient aspect of using numerically simulated data is that the electric field is directly available and can serve as the ground truth. To this end, we apply PC-FDON to a surface dielectric barrier discharge (SDBD) system driven by a DC power supply at a high voltage (HV) of 5~kV. The SDBD comprises a thin brass disk as the HV anode (diameter 20~mm), a quartz dielectric layer, and a hollow alumina cylinder as the cathode (diameter 60~mm). Two electrode configurations are examined: a circular HV electrode and a geared HV electrode with 2~mm-long tips. The corresponding electrostatic fields are simulated using COMSOL Multiphysics. The reader is referred to Appendix~\ref{sec:append_setup} for further details. The simulated field profiles serve as the ground truth ($E_{\mathrm{T}}$) and are used to generate synthetic EFISH profiles $P^{(2\omega)}$ via Eq.~\eqref{eq:3-EFISH_norm}. These EFISH profiles are obtained based on the optical parameters of our previous experiments~\citep{yangInterpretableOperatorlearningModel2026} and detailed in Fig.~\ref{fig:fig6_Expt_setup} of Appendix~\ref{sec:append_setup}. Briefly, a 1064~nm probe beam is focused by a 25~cm lens (for both horizontal and vertical polarization conditions), yielding a Rayleigh length $z_{\mathrm{R}} = 1.35$~mm and a phase-mismatch parameter $u = -0.068$. To provide better correspondence with possible future experiments, the electric field is probed at a distance of 1.5~mm from the lower edge of the HV anode and 1.5~mm from the dielectric surface to ensure that the focused beam does not impinge on either boundary. The generated EFISH profiles are then fed into PC-FDON to yield the predicted electric field components $E_{\mathrm{P}} = \{E_{x,\mathrm{P}},\, E_{y,\mathrm{P}}\}$.

A comparison between $E_{\mathrm{T}}$ and $E_{\mathrm{P}}$ is shown in Fig.~\ref{fig:fig4_Sim_dbd}. Both SDBD configurations exhibit clear OOD behavior (Fig.~\ref{fig:fig4_Sim_dbd}b,f,d,h): at a large fraction of the sampling points along the $z$-axis, the pointwise STD $\sigma(z^{\prime})$ averaged over 100 MC dropout passes far exceeds the 99th-percentile threshold of the validation set. This OOD diagnosis is corroborated by the reconstruction itself, which shows appreciable deviations from the ground truth precisely in the spatial regions where the model uncertainty is highest. The result is physically consistent: the SDBD field profiles -- which feature triple-peak EFISH profiles resulting from the electrode geometry -- fall outside the smooth bell-shaped and double-peak EFISH function families on which PC-FDON was trained, and the uncertainty quantification framework correctly identifies this mismatch.

\subsubsection{Experimental electrostatic fields}~\label{subsubsec: electrostatic_exp}
\begin{figure*}
    \centering
    \includegraphics[width=1\linewidth]{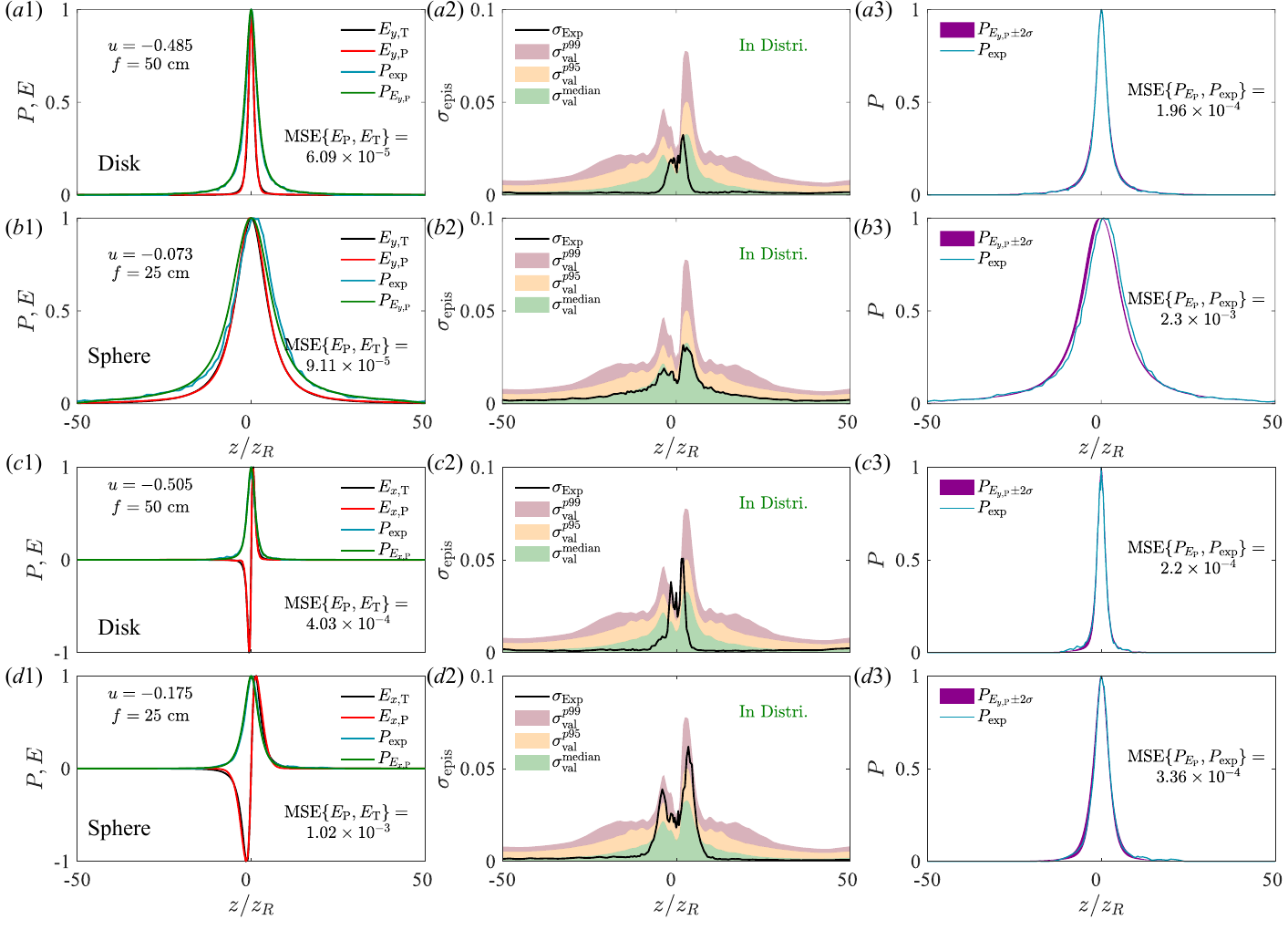}
    \caption{Experimental electrostatic field reconstruction via PC-FDON for (a,\,c)~disk--disk and (b,\,d)~sphere--sphere electrode configurations. Panels (a,\,b) show the reconstructed vertical field profile $E_{y}^{\prime}$ and panels (c,\,d) show the horizontal field profile $E_{x}^{\prime}$, both benchmarked against simulations. The focusing lens focal lengths are $f = 50$~cm (disk--disk) and $f = 25$~cm (sphere--sphere). Sub-panels within each case show (2)~out-of-distribution (OOD) detection results via the exceedance fraction $\mathcal{F}^{(p)}$ and (3)~epistemic uncertainty forwarded into the EFISH signal space.}
    \label{fig:fig5_Expt_electrostatic}
\end{figure*}

Beyond the numerically simulated cases, we apply PC-FDON to experimentally acquired EFISH measurements of electrostatic fields under varying polarization states, optical parameters, and electrode geometries. Two electrode configurations are examined: (\romannumeral1)~disk--disk (5~mm-diameter) and (\romannumeral2)~sphere--sphere (10~mm-diameter), with both geometries separated by a 10~mm vertical inter-electrode gap. For the vertical polarization measurements, the beam is probed 5~mm below the HV electrode; for the horizontal polarization measurements, the probing distance is reduced to 3~mm below the HV electrode to achieve a stronger EFISH signal. It should be noted that the horizontal field component is often much weaker than that of its vertical counterpart, and therefore efforts to increase its magnitude were necessary.

The probe beam wavelength ($1064$~nm) is identical to that of our previous studies~\citep{yangDeepLearningApproach2025a,yangInterpretableOperatorlearningModel2026}, and the optical setup is detailed in Appendix~\ref{sec:append_setup}. Two focusing lenses ($f = 25$~cm and $50$~cm) are used to access different Rayleigh lengths and thus different $u$. As the experiments are carried out at room temperature and atmospheric pressure, the wave-vector mismatch is $\Delta k = -0.483~\mathrm{cm}^{-1}$. Notably, because the Rayleigh length $z_{\mathrm{R}}$ critically affects $u$, a beam profiler is used to estimate $z_{\mathrm{R}}$ for each lens configuration, yielding $z_{\mathrm{R}} \in [1.46, 10.12]$~mm and correspondingly $u \in [-0.175, -0.073]$. The probe beam is polarized at either $0^{\circ}$ or $90^{\circ}$ (i.e., horizontal or vertical, and parallel to the field polarization being sampled) before entering the electric field interaction region. A DC voltage of 5~kV is applied to the HV electrode for the vertical polarization experiments and 16~kV for the horizontal polarization experiments. For all experiments, the data sampling range obeys the sampling criterion as discussed in~\S\ref{subsec:results_polar} before and also in~\citet{yangInterpretableOperatorlearningModel2026}, where the non-sampled region is filled with zero values. Additional details of the experimental setup can be found in Appendix~\ref{sec:append_setup}.

As shown in Fig.~\ref{fig:fig5_Expt_electrostatic}(a1-d1), the PC-FDON predictions for both electrode geometries track closely the ground truth (from simulations), with markedly low reconstruction errors across all tested conditions. The model faithfully captures both the bell-shaped vertical field profiles and the bipolar horizontal field profiles for different Rayleigh lengths and electrode geometries. The uncertainty analysis (Fig.~\ref{fig:fig5_Expt_electrostatic}a2-d2) confirms that all experimental cases fall within the in-distribution regime according to the exceedance criterion of Eq.~\eqref{eq:confidence-levels}, consistent with the high reconstruction accuracy observed. To further assess the physical consistency of the uncertainty estimates, the epistemic model uncertainty is propagated into the EFISH signal space by evaluating the forward integral of Eq.~\eqref{eq:3-EFISH_norm}. The resulting uncertainty bands in the EFISH domain are narrow, corroborating the low reconstruction errors and confirming that the model's confidence is well-placed for these experimental conditions.

Notably, the uncertainty reference baseline used here is derived from the noise-free validation dataset, which represents a relatively conservative criterion. The model is trained with a jitter layer that exposes it to noise levels up to 5\% of the signal amplitude ($\mathrm{SNR} \approx 26$~dB); the MC-dropout-derived uncertainty therefore reflects the model's confidence given the noise input it receives.In plasma experiments, a ground-truth electric field is generally unavailable. In our previous work, we proposed a practical validation approach based on comparing the experimentally acquired EFISH profile with the forward EFISH profile reconstructed from the model's predicted field (Fig.~\ref{fig:fig5_Expt_electrostatic}a3--d3), quantified via MSE or other error metrics. Here, we offer a complementary framework for assessing the reliability of such comparisons, grounded in the MC dropout framework.

When the uncertainty analysis classifies an input as in-distribution, it indicates that the signal lies well within the model's training range or demonstrated robustness envelope, regardless of the noise level present. In this regime, the forward comparison is meaningful: the reconstructed EFISH signal is reliable, and by extension so is the predicted electric field. This is evident in Fig.~\ref{fig:fig5_Expt_electrostatic}b3, where the offset in the EFISH signal is faithfully captured and the field prediction remains accurate. When the uncertainty analysis flags an input as OOD, the interpretation requires further examination, because an OOD classification can arise from two distinct causes: (i)~the underlying electric field profile genuinely falls outside the function families represented in the training set, or (ii)~the experimental noise far exceeds the model's robustness range (i.e., the jitter-layer threshold of $\mathrm{SNR} \approx 26$~dB or the noise-robustness test level of $\mathrm{SNR} = 20$~dB). The former scenario is consistent with the SDBD simulation results (Fig.~\ref{fig:fig4_Sim_dbd}), where the sharp field gradients lie outside the smooth training families and are correctly flagged. The latter scenario indicates that the input carries noise that is beyond the model's learned tolerance, reducing prediction reliability even if the underlying field shape is nominally in-distribution.

To disambiguate these two causes, we recommend the following procedure. First, the noise level should be estimated -- for example, through repeated measurements -- and compared against the training jitter threshold and the validated robustness level. If the noise exceeds these thresholds, it is advisable to reduce the noise to a feasible level before repeating the uncertainty analysis on the denoised input. If noise reduction is not experimentally feasible, the model may have to be retrained with elevated noise levels or additional noise sources (e.g., Poisson noise) to extend its robustness envelope. Once noise-induced OOD behavior has been addressed, the framework can more reliably identify true OOD cases arising from genuinely unseen physical conditions. Overall, this framework provides the experimentalist with a structured, two-level evaluation scheme: the MC dropout uncertainty first assesses whether a given measurement can be trusted, and the forward EFISH comparison then quantifies the accuracy of the resulting reconstruction -- together maintaining a self-consistent validation pathway that continues to avoid the need for a ground-truth electric field.

\subsubsection{Limitations}~\label{subsubsec:limitation}
As has been the practice for our previous works, we discuss here several limitations of the current model that may impact the field reconstruction accuracy. The first few issues relate to the inversion approach used, and are therefore common to our previous models, while the remaining few are unique to the new ML model.

PC-FDON addresses the inverse problem between $P^{(2\omega)}(z_{\mathrm{o}}^{\prime})$ and $E_{\mathrm{ext}}^{\prime}(z^{\prime})$ as described by Eq.~\eqref{eq:3-EFISH_norm} for both polarization states. In constructing the model, several quantities -- the gas hyperpolarizability ($\alpha^{(3)}$), the neutral number density ($N$), and the wave-vector mismatch ($\Delta k$) -- are assumed to be spatially uniform along the beam propagation axis. Although training with multiple phase-mismatch values $u$ allows the model to handle different, but spatially uniform, interaction parameters, non-uniform distributions of these quantities along the $z$-axis (e.g., those arising from steep temperature gradients in combustion) fall outside the current model's scope and would require further development.

Also, because the reconstruction relies on an EFISH profile built up from spatially translated measurements rather than a single-shot acquisition, the approach is inherently limited to reproducible or quasi-steady discharges; stochastic field distributions varying from shot-to-shot cannot be captured with the current approach.

Furthermore, the model currently accepts a single polarization component per inference and supports only two discrete polarization states ($\psi \in \{0,\, 1\}$), meaning that two separate inferences (i.e., the model has to be run twice) are required to reconstruct the full vectorial information of the electric field.
Future work could consider accepting an EFISH signal under various probe beam polarization angles or to jointly input the orthogonal polarization profiles simultaneously, enabling vectorial field reconstruction from a single model inference.

Finally, due to the nature of the training dataset employed, the model continues to be limited to profiles of certain shapes. In particular, bell-shaped and double-peak profiles with an existing axis of symmetry for the vertical field component, and bipolar (or antisymmetric) profiles when probing the horizontal field.

\section{Conclusions}~\label{sec:conclusion}
% ============================================================================
In this work, we introduce the PC-FDON (Polarization-Conditioned Fourier-enhanced Deep Operator Network), a unified architecture for reconstructing spatially-resolved electric field profiles from EFISH measurements acquired under arbitrary field polarization states and optical interaction parameters. Building on our earlier DDON inverse solver, PC-FDON addresses several critical limitations of the prior framework through targeted architectural innovations.

The first innovation is a Fourier-enhanced branch network that performs learned spectral convolutions on the input EFISH signal. By operating in the frequency domain, this branch provides a natural inductive bias for the Gouy-phase and phase-mismatch kernels that govern the EFISH forward model, enabling efficient extraction of multi-scale spectral features without explicit mathematical derivation. The second is a polarization-conditioning mechanism based on Feature-wise Linear Modulation (FiLM) and gated units, whereby a lightweight auxiliary branch encodes the signal polarization label and modulates the `spatial+spectral' features extracted by the EFISH branch. This mechanism enables a single \textit{unified} model to learn the distinct forward mappings associated with vertical and horizontal polarization states, eliminating the need for separate polarization-specific models. The third is a physics-informed loss that enforces self-consistency with the theoretical EFISH (forward) equation during training: the predicted electric field is re-inserted into the forward integral, and the resulting mismatch with the measured EFISH signal is penalized alongside the data-fidelity and classification objectives. This soft physical constraint regularizes the learned operator with the aim of improving reconstruction accuracy.

Indeed, the model achieves good reconstruction accuracies under noise-free, incomplete, and noisy conditions, respectively. Generalization tests on out-of-training-distribution function families confirm that the model retains strong predictive performance on unseen profile shapes, with an accuracy comparable to our previous polarization-specific DDON baseline. A Monte Carlo dropout framework is integrated to provide pointwise epistemic uncertainty estimates without modifying the network architecture. A location-dependent exceedance fraction metric, benchmarked against the validation set, enables systematic out-of-distribution detection (OOD) at the profile level. This uncertainty quantification capability has been demonstrated on both simulated surface dielectric barrier discharge (SDBD) configurations and experimentally acquired electrostatic fields, with both showing good agreement when examined within our proposed uncertainty framework.

Overall, the polarization-conditioning and Fourier-enhancement strategies introduced here are not specific to EFISH and could be transferred to other line-of-sight-integrated optical diagnostics in plasma science, such as emission spectroscopy. More broadly, the architectural principles demonstrated in this work -- spatial and spectral inductive bias via ResNet and Fourier layers, conditional modulation via FiLM, and calibrated uncertainty via MC dropout -- offer a general template for physics-informed operator learning in inverse problems where polarization, phase, or other discrete experimental parameters alter the forward mapping.

\section*{Acknowledgment}
We wish to acknowledge financial support from a Singapore Ministry of Education (MOE) Academic Research Fund Tier 1 Grant (22-5447-A0001), and the NUS HPC Call-for-Projects fund (CFP04-CF-018) for providing the computational resources \& high-performance computing (HPC) facilities. We also gratefully acknowledge an MOE PhD research scholarship for Mr Edwin Setiadi Sugeng and the China Scholarship Council (Grant No. 202406280475) for Miss Yaqi Zhang.

\appendix
\renewcommand\thefigure{A.\arabic{figure}}
\renewcommand\thetable{A.\arabic{table}}
\renewcommand\theequation{A.\arabic{equation}}
\setcounter{figure}{0}
\setcounter{table}{0}
\setcounter{equation}{0}
\section{Hyperparameters of PC-FDON}~\label{sec:append_model}
In this Appendix, we list details of the hyperparameters of the PC-FDON with the best performance (see Table~\ref{tab:PC-FDON_configuration}). All hyperparameters are selected after several rounds of parametric optimization.

\begin{table}
    \caption{Detailed configuration of the PC-FDON. The overall structure is as depicted in Fig.~\ref{fig:fig1_PC-FDON}. For Residual blocks, the stride is 2 with zero padding employed simultaneously. Dropout rate is $0.1$ and the activation function is GELU unless otherwise noted (e.g., Tanh and ReLU). The operator $\otimes$ denotes a dot product and $\oplus$ denotes a concatenation operation.}
    \label{tab:PC-FDON_configuration}
\centering
    \begin{tabular}{l l l c c}
    \toprule
    \textbf{Subnet} & \textbf{Components} & \textbf{Layer sequence} & \textbf{Num} & \textbf{Output shape}\\
    \midrule
    \multirow{5}{*}{Polarization branch}
            & Input layer \{$\psi$\}                  & $[*,109,1]$   & 1 & \\
            & Encoder dense layer ($\mathbf{d}_\psi$, Tanh)                  & $[*,\{109\rightarrow 512\}]\times 1$ & 2 & $[*,512]$\\
            \cline{3-4}
            & \multirow{2}{*}{FiLM layers}
                                            & $[*,512] \times 1$ ($\gamma$ layer)                & 1 & $[*,512]$\\
            &                               & $[*,512] \times 1$ ($\beta$ layer)                & 1 & $[*,512]$\\
    \midrule
    \multirow{10}{*}{EFISH branch}
            & Input layer \{$P_\mathrm{norm}^{(2\omega)},u^\prime$\}                 & $[*,2, 109]$   & 1 & \\
            & Jitter                                & $[*,2, 109]$   & 1 & $[*,2, 109]$\\
            \cline{2-5}
            & \textit{ResNet path} (spatial, $\mathbf{h}_\mathrm{spatial}$): &             &               &\\
            & \multirow{3}{*}{ResNet block}
                                        & $[*,109] \times 32$  & \multirow{4}{*}{3} & \multirow{7}{*}{$[*,512]$}\\
            &                            & $[*,55] \times 64$  &  & \\
            &                            & $[*,28]\times 128$ & & \\
            &                            & $[*,14] \times 256$ & & \\
            \cline{3-4}
            & Adaptive avg pool               & $[*,16] \times 256$ & 1\\
            & Flatten + Dense layer $\mathbf{h}_\mathrm{spatial}$                   & $[*,\{4096\rightarrow 512\}]\times 1$ & 3\\
            \cline{2-5}%%%%%%%%%%%%%%%%%%%%%%%%%%%%%
            & \textit{FNO path} (spectral, $\mathbf{h}_\mathrm{fft}$): &             &               &\\
            & Interpolation                 & Non-uniform $\rightarrow$ uniform gird &               &\\
            & Fourier block                  & \textit{FFT $\rightarrow$ spectral conv $\rightarrow$ iFFT} &     & \\
            &                                & \quad modes:[256,128,32,16]  & 4 &  $[*,2,512]$\\
            & Flatten + Dense layer $\mathbf{h}_\mathrm{fft}$                  & $[*,\{1024\rightarrow 512\}]\times 1$ & 4\\
             \cline{2-5}%%%%%%%%%%%%%%%%%%%%%%%%%%%%%
            & \multirow{8}{*}{Latent space}  & \textit{Dense layer} $\mathbf{h}_{P,u}$:               &    &   \\
            &                                &\quad ResNet $\mathbf{h}_\mathrm{spatial}$ $\oplus$ Fourier $\mathbf{h}_\mathrm{fft}$ & 1 &   $[*,1024]$\\
            &                                &\quad $[*,\{1024\rightarrow 512\}]\times 1$   & 3 & $[*,512]$\\
            &                                & \textit{FiLM layer} $\widetilde{\mathbf{h}}_{P,u}$: $\gamma \otimes \mathbf{h}_{P,u} + \beta$                               &         & $[*,512]$\\
            &                                & Concatenation with $\mathbf{d}_\psi$          &   1    & $[*,1024]$      \\
            &                                & \textit{Gate unit} (Tanh): & & \\
            &                                & \quad Dense $\mathbf{g}_{P,u,\psi}$: $[*,\{1024\rightarrow 512\}]$  & \multirow{2}{*}{3}  & \multirow{2}{*}{$[*,512]$}\\
            &                                & \quad $\widetilde{\mathbf{H}}_{P,u} \leftarrow \widetilde{\mathbf{h}}_{P,u} \otimes \mathbf{g}_{P,u,\psi}$ & & \\
    \midrule
    \textit{\textbf{Classifier}} (ouput 1)
            & Dense layer (ReLu)                      &$[*,\{512\rightarrow 1\}]\times 1$ & 4 &   $[*,1]$\\
    \midrule
    \multirow{7}{*}{Trunk net}
            & Input layer \{$z_\mathrm{o}^\prime$\}                & $[*,1,109]$   & 1 & \multirow{7}{*}{$[*,512]$}\\
            & Position encoder ($\mathbf{z}$, Tanh)        & $[*,109,\{1\rightarrow 128\}]$ & 3 & $[*,109,128]$\\
            \cline{3-4}
            & \multirow{3}{*}{ResNet block}
                                        & $[*,109] \times 32$  & \multirow{3}{*}{2} & \\
            &                            & $[*,55] \times 64$  &                     & \\
            &                            & $[*,28]\times 128$ &                     & \\
            \cline{3-4}
            & Adaptive avg pool               & $[*,16] \times 128$ & 1 & \\
            & Flatten $+$ Dense layer $\boldsymbol{\tau}$                   & $[*,2048 \rightarrow 512]\times 1$ & 3 & \\
    \midrule
    Latent space
            & Branch $\widetilde{\mathbf{H}}_{P,u}$ $\otimes$ Trunk $\boldsymbol{\tau}$ & $[*,\{512 \rightarrow 512\}]$   & 1 & $[*, 512]$\\
    \cline{2-5}
    \multirow{2}{*}{Decoder}
            & Broadcast to query points     & $[*,512] \rightarrow [*,109,512]$ & 1 & $[*, 109, 512]$\\
            & Concatenation with $\mathbf{z}$                & $[*,109,512] \oplus [*,109,128]$ & 1 & $[*, 109, 640]$\\
    \midrule
    \textit{\textbf{Predictor}} (output 2)
            & Point-wise dense layer              & $[*,109,\{640 \rightarrow 1\}]$ & 4 & $[*, 109]$\\

    \bottomrule
    \multicolumn{2}{l}{\textbf{Hyperparameter setting}} & \multicolumn{2}{l}{\textbf{Value}}\\
    \midrule
    \multicolumn{2}{l}{Learning rate (decaying)} & \multicolumn{2}{l}{$1\times 10^{-3}$ (initial)}\\
    \multicolumn{2}{l}{Batch size} & \multicolumn{2}{l}{512}\\
    \multicolumn{2}{l}{Dropout} & \multicolumn{2}{l}{0.1 (default); 0.05 (predictor layer)}\\
    \multicolumn{2}{l}{FNO grid size} & \multicolumn{2}{l}{512}\\
    \multicolumn{2}{l}{PINN $\lambda_{\text{PINN}}$, Classification $\lambda_{\text{CLS}}$} & \multicolumn{2}{l}{$\lambda_{\text{PINN}}=0.05,\,\lambda_{\text{CLS}}=0.5$}\\
    \multicolumn{2}{l}{Epochs} & \multicolumn{2}{l}{200 (early stopping)}\\
    \bottomrule
    \end{tabular}
\end{table}

\renewcommand\thefigure{B.\arabic{figure}}
\renewcommand\thetable{B.\arabic{table}}
\renewcommand\theequation{B.\arabic{equation}}
\setcounter{figure}{0}
\setcounter{table}{0}
\setcounter{equation}{0}
\section{Experiment and simulation details}~\label{sec:append_setup}
\begin{figure*}
    \centering
    \includegraphics[width=1\linewidth]{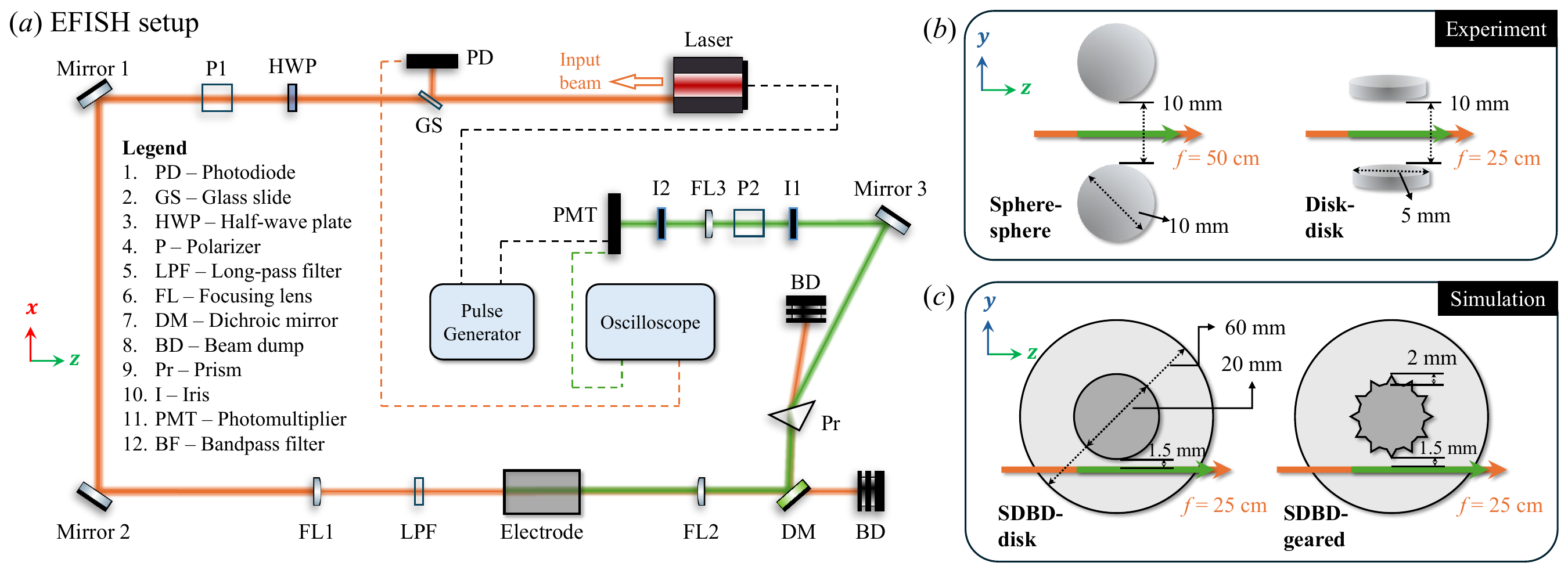}
    \caption{(a)~Schematic of the picosecond EFISH experimental setup. (b)~Isometric and side views of the electrode geometries for the sphere--sphere and disk--disk configurations used in the electrostatic experiments. (c)~Surface dielectric barrier discharge (SDBD) electrode configurations used to generate the simulated electrostatic field profiles and synthetic EFISH profiles.}
    \label{fig:fig6_Expt_setup}
\end{figure*}
\begin{table}
    \centering
    \caption{Details for the electrostatic experiments in \S~\ref{subsubsec: electrostatic_exp} and SDBD simulations in \S~\ref{subsubsec: SDBD}. HV is short for `high voltage'.}
    \begin{tabular}{lccccc}
    \toprule
         Electrode type & Focusing lens & DC Voltage & Polarization & $z_\mathrm{R}$ & $\Delta k$ \\
         \midrule
         \multirow{2}{*}{Disk-disk (Dia. $5$~mm, gap $10$~mm)}
                        & $50$~cm         & $5$~kV          & vertical      & $9.72$~mm       & $-0.483~\mathrm{cm}^{-1}$\\
                        & $50$~cm         & $16$~kV         & horizontal    & $10.12$~mm      & $-0.483~\mathrm{cm}^{-1}$\\
        \cline{2-6}
        \multirow{2}{*}{Sphere-sphere (Dia. $10$~mm, gap $10$~mm)}
                        & $25$~cm         & $5$~kV          & vertical      & $1.46$~mm       & $-0.483~\mathrm{cm}^{-1}$\\
                        & $25$~cm         & $16$~kV         & horizontal    & $3.5$~mm      & $-0.483~\mathrm{cm}^{-1}$\\
        \midrule
        SDBD-disk       & $25$~cm         & $5$~kV          & vertical      & $1.35$~mm       & $-0.5~\mathrm{cm}^{-1}$\\
        (HV electrode: Dia. $20$~mm)      & $25$~cm         & $5$~kV          & horizontal    & $1.35$~mm      & $-0.5~\mathrm{cm}^{-1}$\\
        \cline{2-6}
        SDBD-gear         & $25$~cm         & $5$~kV          & vertical      & $1.35$~mm       & $-0.5~\mathrm{cm}^{-1}$\\
        (HV electrode: Dia. $20$~mm, tip $2$~mm) & $25$~cm         & $5$~kV          & horizontal    & $1.35$~mm      & $-0.5~\mathrm{cm}^{-1}$\\
    \bottomrule
    \end{tabular}
    \label{tab:exp_info}
\end{table}
This appendix provides details of the picosecond EFISH setup used for the electrostatic experiments, as illustrated in Fig.~\ref{fig:fig6_Expt_setup}. All experiments and simulations are conducted at atmospheric pressure and ambient temperature (nominally 1~bar and 25~$^{\circ}$C).

A polarized (vertical or horizontal) 1064~nm fundamental output beam of a $10$~Hz, $3$~mJ, 26~ps Nd:YAG laser (EKSPLA PL2251) is directed through an electrostatic field sustained between a pair of electrodes. Two electrode configurations are used -- sphere-sphere and disk-disk -- whose specifications are listed in Table~\ref{tab:exp_info}. In detail (Fig.~\ref{fig:fig6_Expt_setup}), the pulse energy is monitored via a photodiode with a $1$~ns rise time (Thorlabs DET10A2) at the beam output. The beam polarization is rotated using a $1064$~nm half-wave plate (Thorlabs WPH10M-1064), followed by a beamsplitting polarizer (Thorlabs CCM1-PBS25-1064-HP/M) to select the desired polarization (vertical or horizontal). Thereafter, the beam is focused using either a $25$~cm or $50$~cm focal-length plano-convex lens, and a longpass filter (Thorlabs FGL850S) is positioned immediately after the focusing lens to block stray second-harmonic light generated upstream of the interaction region. These two focal lengths give us different Rayleigh lengths $z_{\mathrm{R}}$, which are approximated by a beam profiler (Thorlabs BC207VIS/M) before each set of experiments. The wave-vector mismatch is constant at $\Delta k = -0.483~\mathrm{cm}^{-1}$, thus resulting in different phase-mismatch values $u$ (see Table~\ref{tab:exp_info}). After passing through the electrostatic field, the beam is collimated by a second plano-convex lens of the same focal length, and the EFISH signal at $532$~nm is separated from the fundamental using a dichroic mirror (HBSY22) and a dispersive prism. The second-harmonic beam is then passed through an additional beamsplitting polarizer (Thorlabs CCM1-PBS25-532-HP/M) to isolate the desired polarization component and is focused by a 10~cm focal-length plano-convex lens into a gated photomultiplier tube (PMT) module with a 1~ns rise time (Hamamatsu H11526-NF). An iris is placed at the PMT entrance for stray-light rejection. Both the photodiode and PMT signals are recorded on a digital oscilloscope (Teledyne LeCroy T3DSO31004) operating at a $5$~GHz sampling rate with 1~GHz analog bandwidth.

\subsection*{Electrostatic experiments}
For the electrostatic experiments described in \S\ref{subsubsec: electrostatic_exp}, the electric field is generated using the sphere--sphere and disk--disk aluminum electrode geometries shown in Fig.~\ref{fig:fig6_Expt_setup}b; the electrode dimensions and operating conditions are summarized in Table~\ref{tab:exp_info}. The beam probes the electric field at mid-height between the electrodes ($5$~mm below the HV electrode) for the vertical polarization measurements, while the probing distance is reduced to $3$~mm below the HV electrode for the horizontal polarization measurements to achieve a stronger EFISH signal. A DC voltage of $5$~kV is applied for the vertical case and 16~kV for the horizontal case, supplied by a 20~kV, 700~$\mu$A DC power supply (Glassman High Voltage, Inc.). Both electrode geometries are translated along the $z$-axis on an optical rail in increments of $\Delta z$ to construct the spatially-resolved EFISH signal profile. The step size is varied between $\Delta z = 1$~mm, $2$~mm, $5$~mm, and $10$~mm depending on the distance from the beam focus, with finer increments used near the focal region. At each $z$-location, the raw signals are averaged over 200 laser shots ($20$~s acquisition time) and divided by the square of the laser intensity after background subtraction.

\subsection*{SDBD simulations}
For the SDBD simulations described in \S\ref{subsubsec: SDBD}, the optical configuration is identical except that only the $25$~cm focal-length plano-convex lens is used. The optical interaction parameters for generating the synthetic EFISH follows that of~\citet{yangInterpretableOperatorlearningModel2026}. The electric field profiles along the beam propagation path are extracted from three-dimensional electrode models and corresponding electrostatic field simulations via COMSOL Multiphysics. Two SDBD electrode configurations are modeled, see Fig.~\ref{fig:fig6_Expt_setup}c: a disk-shaped (smooth) HV electrode and a gear-shaped HV electrode. The disk-shaped configuration is identical to that studied by~\citet{zhangInhibitionPromotionQuasiuniform2026}, with the exception that the dielectric material is quartz rather than PTFE. The gear-shaped configuration shares the same dimensions and materials as the disk-shaped one but features radially protruding teeth of $2$~mm in length uniformly distributed around the circumference of the HV electrode. Both HV electrodes are copper and 0.3~mm thick. The quartz dielectric layer has a diameter of $60$~mm and a thickness of $0.3$~mm, while the aluminum ground electrode has the same diameter as the dielectric layer and a thickness of $0.05$~mm. Space charge density in the air gap and surface charge deposited on the dielectric are neglected, and the electrostatic field is obtained by solving the Laplace equation via the AC/DC module in the electrode interface of COMSOL:
\begin{equation}
    \nabla \cdot \left(\varepsilon_0 \, \varepsilon_\mathrm{r} \, \nabla\phi \right) = 0, \qquad
    \mathbf{E} = -\nabla\phi,
    \label{eq:laplace}
\end{equation}
where $\mathbf{E}$ is the electric field vector, $\phi$ is the electric potential, and $\varepsilon_{0}$ and $\varepsilon_{\mathrm{r}}$ denote the vacuum permittivity and the relative permittivity of each medium, respectively. Dirichlet boundary conditions are applied on the electrode surfaces, with the electric potentials of the HV and ground electrodes prescribed as $5$~kV and $0$~V, respectively. At the air--quartz interface, continuity of the electric potential and the normal component of the electric displacement field is imposed. A homogeneous Neumann boundary condition is applied on the outer boundaries of the computational domain. The maximum element size is set to $1$~mm, with local mesh refinement applied to the electrode tips and their surrounding regions (minimum element size $0.003$~mm) to resolve the steep electric field gradients near the tooth tips. To generate the synthetic EFISH profiles under both polarization states, the beam is positioned $1.5$~mm from the lower edge of the HV anode and $1.5$~mm above the dielectric surface, ensuring that the focused beam does not impinge on either boundary. As in the electrostatic experiments, the electrodes are translated along the $z$-axis to yield the spatially resolved EFISH signal profile.

\renewcommand\thefigure{C.\arabic{figure}}
\renewcommand\thetable{C.\arabic{table}}
\renewcommand\theequation{C.\arabic{equation}}
\setcounter{figure}{0}
\setcounter{table}{0}
\setcounter{equation}{0}
\section{Optical interaction parameters}~\label{sec:append_optical_para}
In this appendix, we present an overview of several optical configurations commonly employed in typical EFISH experiments (see Table.~\ref{tab:optical_para}). These setups represent the most widely adopted arrangements in the field, and their operational parameters are used to justify the range covered by our training dataset and that which can be handled by our PC-FDON model.

\begin{table}
    \centering
    \caption{Range of optical interaction parameters commonly used in EFISH experiments. Beam quality is set at $M^2=1.1$. SH denotes the generated second-harmonic light.}
    \begin{tabular}{lccccc}
    \toprule
         Laser wavelength $\lambda^{(\omega)}$ & Lens $f$ & Temp. & $z_\mathrm{R}$ & $\Delta k$ & $u$ \\
         \midrule
         \multirow{6}{*}{$532$~nm (SH $\lambda^{(2\omega)}$: $256$~nm)}
                        & \multirow{2}{*}{$10$~cm}         & $25~^\circ\mathrm{C}$          &  $0.23$~mm      & $-4.41~\mathrm{cm}^{-1}$       & $-0.1$ \\
                        &                                  & $200~^\circ\mathrm{C}$         &  $0.23$~mm      & $-2.78~\mathrm{cm}^{-1}$       & $-0.065$ \\
                        \cline{2-6}
                        & \multirow{2}{*}{$25$~cm}         & $25~^\circ\mathrm{C}$          &  $1.46$~mm      & $-4.41~\mathrm{cm}^{-1}$       & $-0.642$ \\
                        &                                  & $200~^\circ\mathrm{C}$         &  $1.46$~mm      & $-2.78~\mathrm{cm}^{-1}$       & $-0.405$ \\
                        \cline{2-6}
                        & \multirow{2}{*}{$50$~cm}         & $25~^\circ\mathrm{C}$         & $5.82$~mm      & $-4.41~\mathrm{cm}^{-1}$       & $-2.57$ \\
                        &                                  & $200~^\circ\mathrm{C}$         & $5.82$~mm      & $-2.78~\mathrm{cm}^{-1}$       & $-1.62$ \\
                        \midrule
        \multirow{6}{*}{$1064$~nm (SH $\lambda^{(2\omega)}$: $532$~nm)}
                        & \multirow{2}{*}{$10$~cm}         & $25~^\circ\mathrm{C}$          &  $0.23$~mm      & $-0.483~\mathrm{cm}^{-1}$       & $-0.011$ \\
                        &                                  & $200~^\circ\mathrm{C}$         &  $0.23$~mm      & $-0.304~\mathrm{cm}^{-1}$       & $-0.007$ \\
                        \cline{2-6}
                        & \multirow{2}{*}{$25$~cm}         & $25~^\circ\mathrm{C}$          &  $1.46$~mm      & $-0.483~\mathrm{cm}^{-1}$       & $-0.071$ \\
                        &                                  & $200~^\circ\mathrm{C}$         &  $1.46$~mm      & $-0.304~\mathrm{cm}^{-1}$       & $-0.044$ \\
                        \cline{2-6}
                        & \multirow{2}{*}{$50$~cm}         & $25~^\circ\mathrm{C}$         & $5.82$~mm      & $-0.483~\mathrm{cm}^{-1}$       & $-0.281$ \\
                        &                                  & $200~^\circ\mathrm{C}$         & $5.82$~mm      & $-0.304~\mathrm{cm}^{-1}$       & $-0.177$ \\
    \bottomrule
    \end{tabular}
    \label{tab:optical_para}
\end{table}

\printcredits

\clearpage

\bibliographystyle{model1-num-names}
%\bibliographystyle{cas-model2-names}

\bibliography{FDON_ref}



\end{document}